\documentclass[iop]{emulateapj}
\usepackage{euscript}
\usepackage{amsmath}
\usepackage{subfigure}
\usepackage{graphicx}

\shorttitle{Effects of Multiple-scale Driving}
\shortauthors{Yoo \&  Cho}

\begin{document}

\title{Effects of multiple-scale driving on turbulence statistics}

\author{Hyunju Yoo\altaffilmark{1} and Jungyeon Cho\altaffilmark{2}}
\affil{Department of Astronomy and Space Science, Chungnam National University, Daejeon, Korea}

\altaffiltext{1}{hyunju527@gmail.com}
\altaffiltext{2}{jcho@cnu.ac.kr; Corresponding author}

\begin{abstract}
Turbulence is ubiquitous in astrophysical fluids such as the interstellar medium (ISM) 
and the intracluster medium (ICM). 
In turbulence studies, it is customary to assume that fluid is driven on a single scale. 
However, in astrophysical fluids, there can be many different driving mechanisms that act 
on different scales. 
If there are multiple energy-injection scales, 
the process of energy cascade and turbulence dynamo will be different 
compared with the case of single energy-injection scale. 
In this work, 
we perform three-dimensional incompressible/compressible magnetohydrodynamic (MHD) turbulence simulations. 
We drive turbulence in Fourier space in two wavenumber ranges, 2$\leq$k$\leq$$\sqrt12$ (large-scale) 
and 15 $\lesssim$k$\lesssim$ 26 (small-scale). 
We inject different amount of energy in each range 
by changing the amplitude of forcing in the range. 
We present the time evolution of the kinetic and magnetic energy densities and discuss 
the turbulence dynamo in the presence of
energy injections at two scales. 
We show how kinetic, magnetic and density spectra are affected 
by the two-scale energy injections and we 
discuss the observational implications. 
In the case $\epsilon_L < \epsilon_S$, where $\epsilon_L$ and $\epsilon_S$ are
energy-injection rates at the large and small scales, respectively, our results show that
even a tiny amount of large-scale energy injection can significantly change the properties
of turbulence.
On the other hand, when $\epsilon_L \gtrsim \epsilon_S$, the small-scale driving does not
influence the turbulence statistics much unless $\epsilon_L \sim \epsilon_S$. 
\end{abstract}


\section{Introduction}

Turbulence is a common phenomenon in astrophysical fluids (see, for example, Elmegreen \& Scalo 2004) and
it is obvious that most astrophysical fluids are permeated by magnetic fields 
(Brandenburg \& Subramanian 2005). 
Such magnetized fluids can be investigated by numerical simulations of driven magnetohydrodynamic 
(MHD) turbulence (e.g. Biskamp 2003).

To maintain turbulence in fluids, energy must be injected into the fluids. 
Many turbulence simulations have been performed with either solenoidal ($\nabla$$\cdot$$\bf f$$=$0) 
or compressive ($\nabla$$\times$$\bf f$$=$0) forcing 
(e.g., Meneguzzi, Frisch, \& Pouquet 1981; Cho \& Vishniac 2000; Brandenburg 2001; 
Ostriker, Stone, \& Gammie 2001;
Federrath et al. 2008). 
Most of these studies have adopted energy injection on a single scale in Fourier space (wavenumber space). 
However, multiple driving scales should be considered in simulations in order to better
imitate real astrophysical fluids, such as the interstellar medium (ISM) and 
the intracluster medium (ICM), 
because multiple astrophysical driving mechanisms may act on different scales simultaneously
in those systems.

There are many possible energy sources for ISM turbulence.
Mac Low (2004) examined available driving mechanisms for turbulence in the ISM. 
Magnetorotational instabilities, gravitational instabilities, protostellar outflows, 
expansion of H II regions, stellar winds from massive stars, and supernova explosions
have been considered as candidates for energy sources of ISM turbulence. 
Haverkorn et al. (2008) observed  Faraday rotation of extragalatic radio sources 
through the Galactic plane and determined the outer scale of turbulence in the Galactic ISM. 
They suggested that stellar sources, such as stellar winds and protostellar outflows, drive turbulence on parsec scales in the spiral arms and
supernova and superbubble explosions on $\sim$100 parsec scales in the interarm regions.
Han et al. (2004) showed the large-scale magnetic energy spectrum in our Galaxy and suggested 
that ISM turbulence is driven by stellar winds and supernova explosions 
on scales from 10 parsecs to 100 parsecs. 

It is also clear that the ICM is in a turbulent state with multiple driving scales. 
Schuecker et al. (2004) obtained a pressure map of the Coma cluster using XMM-Newton data 
and derived properties of turbulence from the map.
According to their result, the largest eddy size and the smallest eddy size in 
 the central region of the cluster are 145 kpc and 20 kpc, respectively. 
Many possible driving mechanisms exist for ICM turbulence. 
First, there are mechanisms that can provide energy injection on large scales.
For example, cosmological shocks (Ryu et al. 2003;
Pfrommer et al. 2006) or mergers (De Young 1992; Tribble
1993; Norman \& Bryan 1999; Roettiger et al. 1999; Ricker \& Sarazin 2001) can
produce turbulence in which the outer scale 
is similar to the size of the entire cluster.
 In fact, the
outer scale of turbulence observed in some simulations during
the formation of galaxy clusters is up to $\sim$ several hundred kpc
(Norman \& Bryan 1999; Ricker \& Sarazin 2001), which is
a few times smaller than the cluster size of $\sim$Mpc.
Second, there are also mechanisms that can provide energy injection on small scales.
For example, infall of small structures (Takizawa 2005), AGN jets (see,
for example, Scannapieco \& Br\"uggen 2008), or galaxy wakes
(Roland 1981; Bregman \& David 1989; Kim 2007) can produce turbulence in which
the outer scale is much smaller than the size
of a cluster.

Due to variety of turbulence driving scales in the ISM and the ICM, 
it is necessary  to inject energy
on several different scales to simulate turbulence in those systems. 
However, there has been no rigorous research in this direction. 
In this paper, we study MHD turbulence driven at two scales.
We mainly focus on the behavior of kinetic, magnetic, and density spectra
in the presence of the driving at two scales.

The outline of this study is as follows.
We start by explaining numerical methods, initial conditions and forcing used for this work in Section 2. 
Theoretical expectations are given in Section 3. 
Then, results for incompressible MHD turbulence are shown  in Section 4. 
Results for compressible MHD turbulence, especially
density, rotation measure (RM), and  velocity centroid (VC) spectra, are provided in Section 5. 
We discuss astrophysical implications in Section 6 and give summary in Section 7.
 
\section{Numerical Methods}
We use both incompressible and compressible MHD codes.
The incompressible MHD code is based on the pseudospectral method
and the compressible code is based on an essentially Non-Oscillatory (ENO) scheme.

\subsection{Incompressible Code} 
\label{sect:code1}
We directly solve the incompressible MHD equations in a periodic box of size 2$\pi$
using a pseudospectral code:

\begin{eqnarray}
{\partial {\bf v} }/{\partial t} = -(\nabla \times {\bf v})\times {\bf v} 
   + (\nabla \times {\bf B})\times {\bf B} \nonumber\\
  + {\nu}{\nabla}^2 {\bf v} + {\bf f} + {\nabla P'},  \\
{\partial {\bf B} }/{\partial t} = \nabla \times ({\bf v} \times{\bf B}) 
   + \eta {\nabla}^2{\bf B} 
\end{eqnarray}
where $\nabla$$\cdot$$\bf v$=$\nabla$$\cdot$$\bf B$$=$0 and $P'$=$P$+$v^2$/2. 
Here, $\bf v$ is velocity, $\bf B$ is the magnetic field divided 
by $\sqrt {4\pi\rho }$,  
$\nu$ is viscosity, $\eta ~(=\nu)$ is magnetic diffusivity, 
$P$ is pressure,  
and $\bf f$  is a random forcing term. 
The magnetic field consists of a uniform background field $\bf B_{0}$  
 and a fluctuating field $\bf b$. 
 
In this work, we drive turbulence in Fourier space and consider only solenoidal ($\nabla \cdot {\bf f}=0$) forcing. 
We use 22 large-scale forcing components with 2 $\leq$k$\leq$$\sqrt12$ 
and 100 small-scale forcing components with 15 $\lesssim$k$\lesssim$ 26, where
$k$ is the wavenumber.
Therefore, peaks of energy injections occur at k $\sim$ 2 ($\equiv$ $k_{L}$) 
and at k $\sim$ 20 ($\equiv$ $k_{S}$). 
    The 22 large-scale forcing components are nearly isotropically
   distributed  in the range 2 $\leq$k$\leq$$\sqrt12$ in Fourier space.
   The wavenumbers used are 2 (three components),
   $\sqrt{6}$ (12 components), 3 (three components), and
   $\sqrt{12}$ (four components).
   The amplitude of
   each large-scale forcing component  fluctuates randomly. On average, the amplitude of
   each component is same.
    Therefore, the average large-scale driving wavenumber will be $k \sim 2.5$.  
     However, since our kinetic spectra clearly show a peak at k=2 
     in the presence of large-scale driving,
     we assume that the peak of large-scale energy injection occurs at k$\sim$2.
    If we take $k_L =2.5$, the coefficient on the right-hand side 
    in Equation (\ref{eq:theo_exp}) will be $\sim 32$.
    For the small-scale driving, we randomly selected 100 Fourier modes in the range $15<k<26$. 
    On average the amplitude of each small-scale forcing mode is same.
 
Incompressible turbulence simulations are performed on a spatial grid of $256^{3}$ points.
The mean magnetic field $B_0$ is either 0.001 
(weakly magnetized cases) or 1.0 (strongly magnetized cases)\footnote{  
     The strength of the mean field is equal to the amplitude of the 
     $k=0$ Fourier mode, which does not change with time.}.
There is either large-scale velocity (weakly magnetized cases) or 
no velocity (strongly magnetized cases) at the beginning of a simulation. 
We consider only cases where the viscosity is equal to the
magnetic diffusivity: $\nu=\eta$.
We use hyper-viscosity of the form $\nu_3(\nabla^2)^3 {\bf v}$ for the viscosity term and
a similar expression for the magnetic diffusion term.
We refer the reader to Cho \& Vishniac (2000) or Cho et al. (2009) for further details of the code.

\subsection{Compressible Code} 
\label{sect:code2}
In order to see the effect of multiple driving scales on compressible turbulence, 
we use an ENO scheme (see Cho $\&$ Lazarian 2002) 
to solve the ideal compressible MHD equations in a periodic box of size $2\pi$ :

\begin{eqnarray}
{\partial \rho    }/{\partial t} + \nabla \cdot (\rho {\bf v}) =0,  \\
{\partial {\bf v} }/{\partial t} + {\bf v}\cdot \nabla {\bf v} 
   +  \rho^{-1}  \nabla(a^2\rho)\nonumber\\
   - (\nabla \times {\bf B})\times {\bf B}/4\pi \rho ={\bf f},  \\
{\partial {\bf B}}/{\partial t} -
     \nabla \times ({\bf v} \times{\bf B}) =0, 
\end{eqnarray}
with $\nabla$$\cdot$$\bf B$$=$0 and an isothermal equation of state $P$=$C^{2}_{s}\rho$, 
where  $C_{s}$ is the sound speed and $\rho$ is density. 
The mean magnetic field $B_0$ is either 0.01 
(weakly magnetized cases) or 1.0 (strongly magnetized cases).
We use $256^3$ grid points.
 Forcing and other setups are the same as those in incompressible simulations. 
In this work, we consider only the cases with $M_s ~(\equiv v_{rms}/C_s)\sim 1$, where
$v_{rms}$ is the r.m.s. velocity, and we do not consider self-gravity, 
cooling effects and radiative energy transfer in order to concentrate on the effects of the two-scale forcing.

\subsection{Notations} 
In this work, we perform simulations   
          with different energy-injection rates on large and small scales,
           which are achieved by adjusting the amplitudes of the forcing.
energy-injection rates $\epsilon$ (= $\bf f$$\cdot$$\bf v$)  
and their large-scale to small-scale ratios $R_\epsilon$ ($\equiv \epsilon_L/\epsilon_S$)
are listed in Tables 1 to 3. 
We use the notation ``$a$\_$Lb$\_$Sc$'', where
$a$ denotes the simulation property,  ``$IH$'' (Incompressible Hydrodynamic), 
``$IW$''
(Incompressible Weakly Magnetized), ``$IS$'' (Incompressible Strongly Magnetized), 
``$CW$'' (Compressible Weakly Magnetized), 
and ``$CS$'' (Compressible Strongly Magnetized), and
$b$ and $c$ refer to the amplitudes of the large-scale and small-scale forcing, respectively.  

\section{Theoretical Expectations}

\subsection{Expected Scaling Relations for Velocity}   \label{sect:3.1}

Suppose that hydrodynamic turbulence is driven at a small scale $l_S$.
According to Kolmogorov's theory (Kolmogorov 1941), energy-injection rate $\epsilon_S$ is 
\begin{equation}
\epsilon_S \sim \frac{v_S^3}{l_S},   %
\end{equation}
where 
the subscript `S' denotes the small scale
and $v_S$ is the velocity dispersion at the scale $l_S$.  
{}From this, we have
\begin{equation}
   v_S \sim ( \epsilon_S l_S )^{1/3}   \label{eq:v_v}
\end{equation} 
and
\begin{equation}
   E(k_S) \sim v_S^2 l_S \sim  \epsilon_S^{2/3} l_S^{5/3},  \label{eq:E_E}
\end{equation} 
where $E(k_S)$ is the 
value of the energy spectrum at $k_S$.

If we inject an additional energy at a larger scale $l_L ~(l_L > l_S)$ with
an energy-injection rate $\epsilon_L$, we will be able to see two peaks in energy spectrum,
one at $k_L ~(\sim 1/l_L)$ and the other at $k_S ~(\sim 1/l_S)$. 
{}From Equations (\ref{eq:v_v}) and (\ref{eq:E_E}), we have
\begin{eqnarray}
   \frac{ v_L }{ v_S } \sim \left( \frac{ \epsilon_L l_L }{ \epsilon_S l_S } \right)^{1/3},  \label{eq:vv} \\
   \frac{ E(k_L) }{ E(k_S) } \sim \left( \frac{ \epsilon_L }{ \epsilon_S } \right)^{2/3}
                                  \left( \frac{ l_L }{ l_S } \right)^{5/3}.   \label{eq:EE}
\end{eqnarray}
If the energy-injection rate at the large scale $\epsilon_L$ is much smaller than that at the small
scale $\epsilon_S$, the spectral peak at $k_L$ will be lower than that at $k_S$: $E(k_L)<E(k_S)$.
Equation (\ref{eq:EE}) implies that 
even a tiny amount of energy injection at $l_L$ will suffice
to make the spectral peak at $k_L$ larger than that at $k_S$:
\begin{equation}
   \frac{ E(k_L)}{E(k_S)} \gtrsim 1, \mbox{~~if~~}   \frac{ \epsilon_L}{\epsilon_S} ~(\equiv R_\epsilon )
   \gtrsim \left( \frac{l_S}{l_L} \right)^{5/2}
        = \left( \frac{k_L}{k_S} \right)^{5/2}.   \label{eq:EE_2}
\end{equation}
Note that $E(k_L) \sim E(k_S)$ does not mean that $v_L \sim v_S$.
Since $v_L/v_S \sim \sqrt{ k_LE(k_L) }/\sqrt{ k_S E(k_S) }$ and $k_L < k_S$, $v_L < v_S$ if $E(k_L)=E(k_S)$.
In order to have $v_L > v_S$, we need a relatively higher, but still small, 
large-scale energy-injection rate (see Equation~(\ref{eq:vv})):
\begin{equation}
   \frac{ v_L }{ v_S } \gtrsim 1, \mbox{~~if~~}    
   \frac{ \epsilon_L}{\epsilon_S}  \gtrsim  \frac{l_S}{l_L} 
        =  \frac{k_L}{k_S}.  \label{eq:vv1}
\end{equation}
If the ratio of energy-injection rates $\epsilon_L/\epsilon_S$ 
is between $(l_S/l_L)^{5/2}$ and $\sim 1$, i.e.~if
\begin{equation}
\left(  \frac{l_S}{l_L} \right)^{5/2} \lesssim \frac{ \epsilon_L}{\epsilon_S} \lesssim 1,
\end{equation}
we will be able to see two spectral peaks, a higher peak at $k_L$ and a lower peak at $k_S$.

If we inject more energy at the large scale than at the small scale, i.e.~if
$\epsilon_L > \epsilon_S$, the peak at $k_S$ will become invisible and we will see
only one peak at $k_L$.
We can show this in the following way. Suppose that there is only one energy injection at $l_L$.
Due to the constancy of the energy cascade rate in the inertial range, 
we can write 

\begin{equation}
\epsilon_{L} \sim \frac{v_L^3}{l_L} = \frac{v_S^{'3}}{l_S}. 
\label{eq:7}
\end{equation}
where $v_S^{'}$ is the velocity fluctuation at $l_S$ produced by the energy cascade from $l_L$.
Now, let us inject additional energy at $l_S$.
The additional energy injection at the small-scale $l_S$ produces its own velocity fluctuation:

\begin{equation}
\epsilon_{S} \sim \frac{v_S^{3}}{l_S},   
\label{eq:8}
\end{equation}
where $v_S^{}$ is the velocity fluctuation induced by the additional driving at the small scale. 
In order for the small-scale peak to be visible, 
$v_S^{}$ should be larger than $v_S^{'}$:
\begin{equation}
v_{S}^{} \gtrsim v_{S}^{'} \rightarrow \left(\epsilon_S l_S\right)^{\frac{1}{3}} \gtrsim
\left(\epsilon_L l_S\right)^{\frac{1}{3}}  .
\label{eq:9}
\end{equation}
Therefore, if $\epsilon_{L} > \epsilon_{S}$, $v_{S}^{'} > v_{S}$ and
it will be difficult to see the peak at $k\sim k_S$ in the kinetic energy spectrum.

\subsection{$E(k_L)/E(k_S)$ in Our Simulations}  
\label{sect:3.2}

In Equation (\ref{eq:EE}),  $k_L$ and $k_S$ are the central wavenumbers of  large- and small-scale forcing, 
respectively. We take 
$k_L\sim2$ and $k_S\sim20$ in our simulations. Therefore,

\begin{equation}
\frac{E(k_{L})}{E(k_{S})}\sim 46\left(\frac{\epsilon_{L}}{\epsilon_{S}}\right)^\frac{2}{3}
\label{eq:theo_exp}
\end{equation}

The relations in this section will be compared with results of hydrodynamic simulations in Section \ref{sect:hd} and 
those of MHD simulations in Section \ref{sect:4}.
Note that the relation in Equation (\ref{eq:EE}) allows us to determine the energy-injection rates from observations.
That is, if we measure the ratio $E(k_L) /E(k_S)$ from observations, we can obtain $\epsilon_L/\epsilon_S$.

\subsection{Numerical Tests for Incompressible Hydrodynamic Turbulence}  
\label{sect:hd}
 To check the scaling relations, 
we perform hydrodynamic simulations. 
We use the pseudospectral code described in
Section \ref{sect:code1} (without magnetic field) 
to solve the incompressible Navier-Stokes equation. 
We start the simulations with zero initial velocity and drive turbulence in two wavenumber ranges, 
$k_L \sim 2$ and $k_S \sim 20$, in
Fourier space (see Section \ref{sect:code1} for details).

\begin{figure*}
  \includegraphics[width=0.32\textwidth]{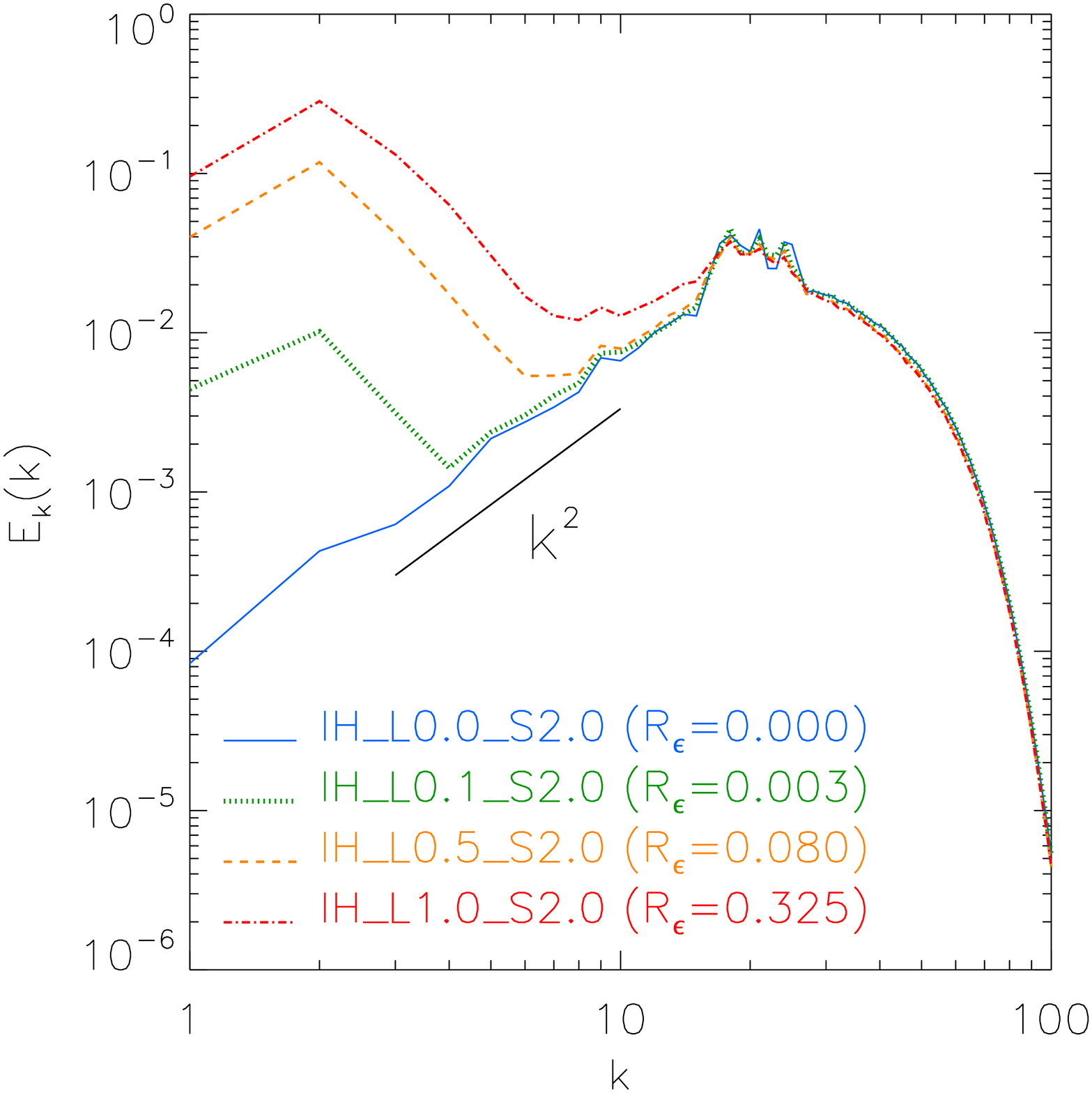}
  \hfill
  \includegraphics[width=0.32\textwidth]{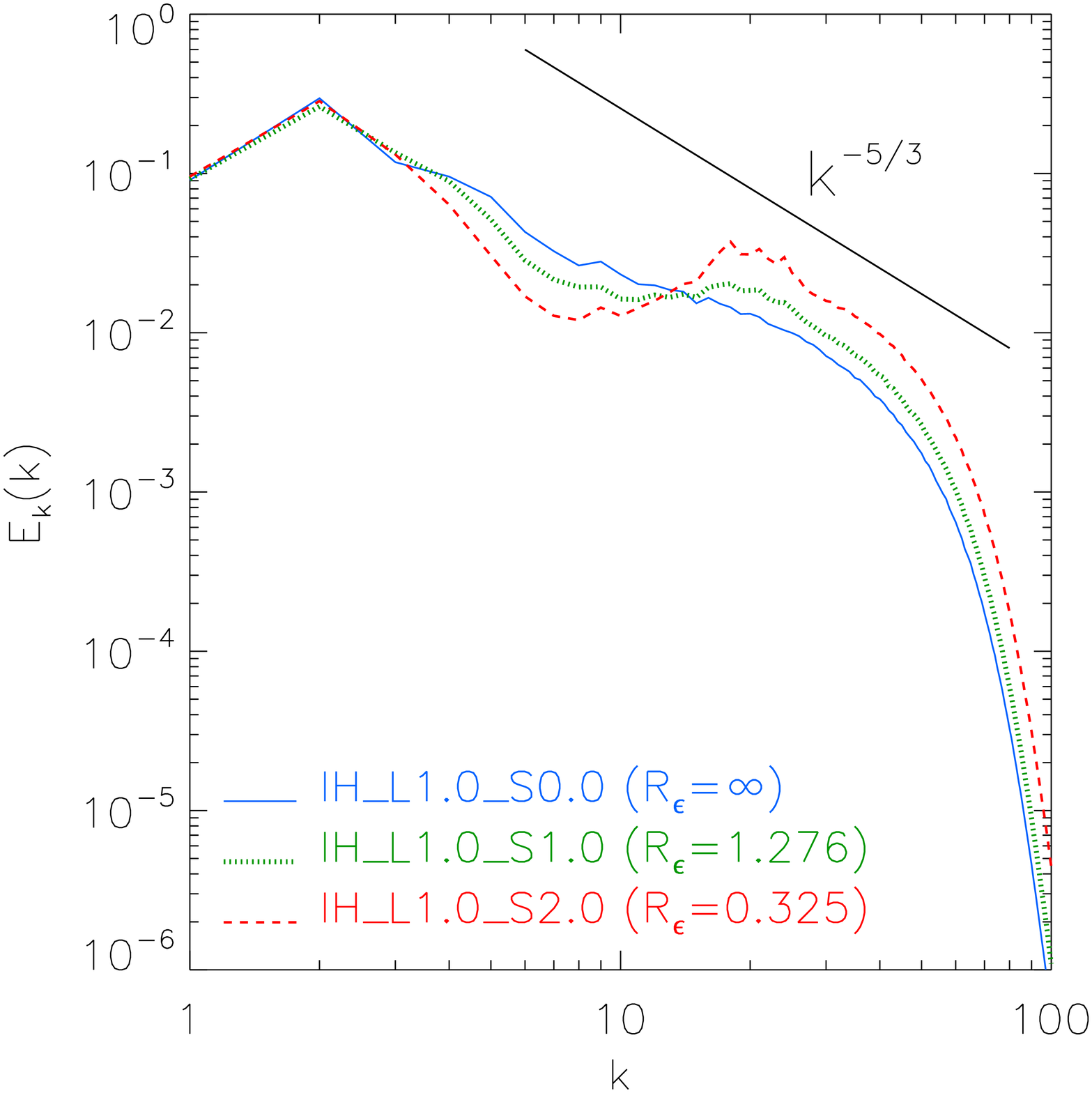} 
  \hfill
  \includegraphics[width=0.32\textwidth]{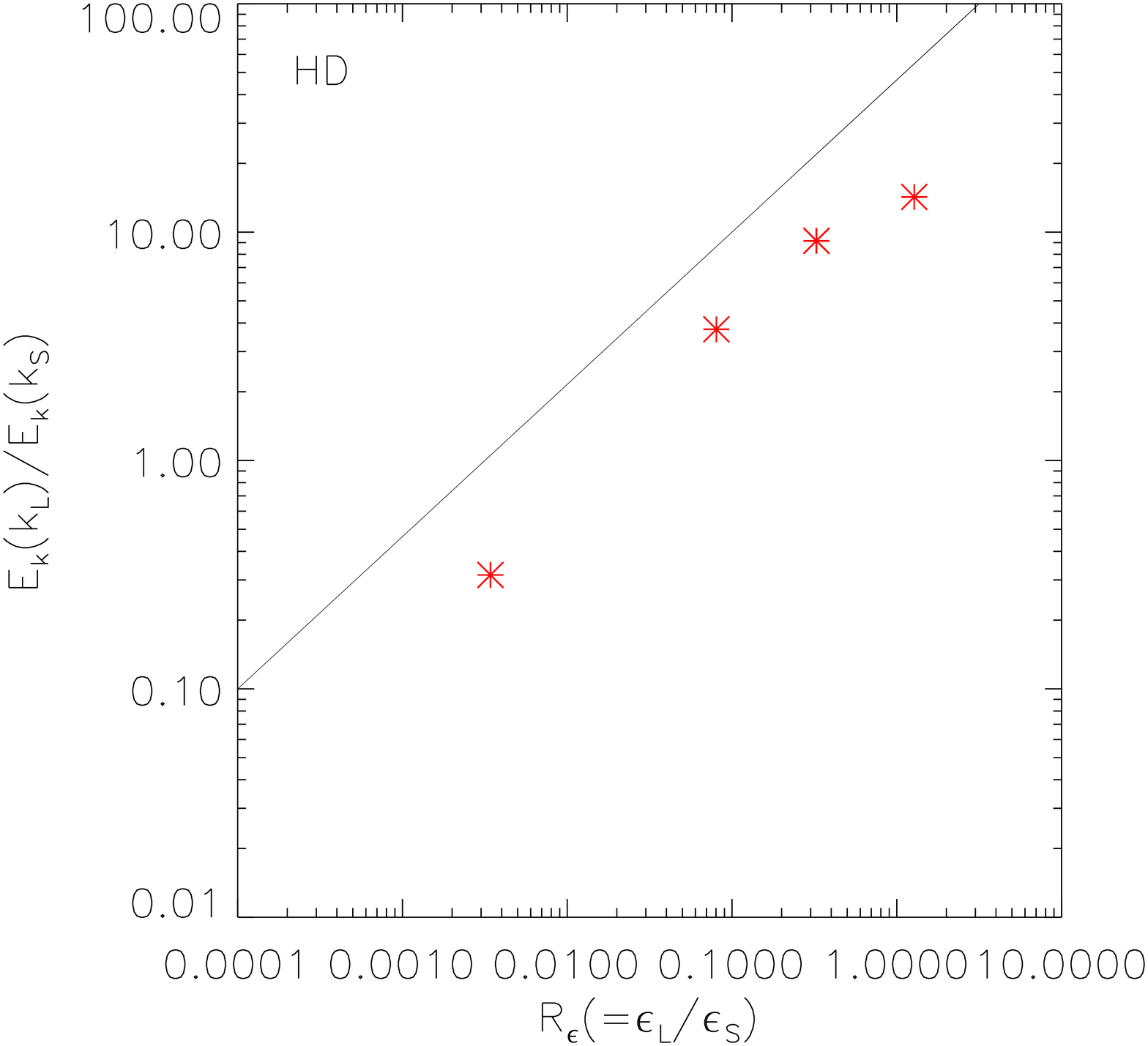}
  \\
  \caption{Hydrodynamic turbulence.
   Left: runs with dominant small-scale driving 
  ($R_\epsilon = \epsilon_L/\epsilon_S <1$).
  We use the same prescribed small-scale driving for all simulations, while 
  the amplitudes of the large-scale driving are different in different runs. 
  In the run with $R_\epsilon =0$, we drive turbulence only at $k_S\sim 20$.
  In other runs, we drive turbulence both at $k_S\sim 20$ and $k_L\sim 2$.
  Middle: runs with $R_\epsilon = \epsilon_L/\epsilon_S \gtrsim 1$. 
  We use the same prescribed large-scale driving for all simulations, while 
  the amplitudes of the small-scale driving are different in different runs.
  In the run with $R_\epsilon =\infty$, we drive turbulence only at $k_L\sim 2$.
  In other runs we drive turbulence both at $k_S\sim 20$ and $k_L\sim 2$.
  Right: the relation between $E(k_L)/E(k_S)$ and $\epsilon_L/\epsilon_S$.
  The black solid line is for the theoretical expectation [Eq. (\ref{eq:theo_exp})].
  Results are calculated at t$\sim$15.}
\label{f:hydro}
\end{figure*}

The left and middle panels of Figure \ref{f:hydro} show kinetic energy spectra for various values 
of $R_\epsilon ~(= \epsilon_L/\epsilon_S)$.
In the simulations shown in the left panel, we fix the amplitude of 
small-scale forcing and vary that of large-scale forcing.
Note that $ \epsilon_L/\epsilon_S \lesssim 1$ for all the simulations in the left panel.
In Run $IH\_L0.0\_S2.0$ (the blue solid line), in which we drive turbulence only at the small scale,
we can see a spectral peak at $k_S \sim 20$ and
the spectral slope for $k>k_S$ is consistent with +2.
In Run $IH\_L0.1\_S2.0$ (the green dotted line), 
in which a small fraction ($R_\epsilon \sim 0.003$) of additional energy is injected
at $k_L \sim 2$, we can see an additional spectral peak at $k_L$.
As the value of $R_\epsilon$ increases, the height of the peak at $k_L$ (i.e.~$E(k_L)$) goes up.
Even if $\epsilon_L \ll \epsilon_S$ (see Run $IH\_L0.5\_S2.0$ and Run $IH\_L1.0\_S2.0$),
the large-scale spectral peak can be higher than the small-scale one.
This is because $k_S \gg k_L$.
In fact, Equation~(\ref{eq:EE}) with $k_L \sim 2$ and $k_S \sim 20$ implies that 
\begin{equation}
   \frac{ E(k_L)}{E(k_S)} \gtrsim 1, \mbox{~~if~~}   \frac{ \epsilon_L}{\epsilon_S} ~(=R_\epsilon) \gtrsim 
         10^{-2.5}.   \label{eq:18}
\end{equation}

The middle panel of Figure \ref{f:hydro} shows the behavior of the kinetic energy spectrum
for $\epsilon_L/\epsilon_S \gtrsim 1$.
For the sake of numerical convenience,
we fix the amplitude of the large-scale forcing 
and vary that of the small-scale forcing
in the simulations shown in the middle panel.
In Run $IH\_L1.0\_S2.0$, where $\epsilon_L /\epsilon_S \approx 0.325$, we can clearly see
two peaks.
In Run $IH\_L1.0\_S1.0$, where $\epsilon_L/\epsilon_S \approx 1.276$, we can still see
two distinct peaks, one at $k_L \sim 2$ and the other at $k_S \sim 20$.
However, the peak at $k_S$ is less pronounced compared with that of Run $IH\_L1.0\_S2.0$.
In Run $IH\_L1.0\_S0.0$, in which we drive turbulence only at the large scale, 
we can see only one peak at $k_L \sim 2$.
The behavior of the spectra in the figure is consistent with
our expectation that two peaks 
in the spectrum become visible when $\epsilon_L/\epsilon_S \lesssim 1$.

The right panel of Figure \ref{f:hydro} shows that the measured 
$R_\epsilon$ - $E(k_{L})/E(k_{S})$ relation
is consistent with the theoretical expectation in Equation (\ref{eq:theo_exp}).
The values of $E(k_{L})/E(k_{S})$ are actually slightly lower than
the expectation (the solid guide line in the right panel), which might be due to
the uncertainty in determining $k_L$.

\section{Velocity and Magnetic Fields in Incompressible MHD Turbulence}  
\label{sect:4}
In this section, we use the incompressible MHD code (see Section \ref{sect:code1}).
It is known that astrophysical fluids are compressible, 
but we first perform incompressible turbulence simulations to focus on the effects of driving 
without any compression effect. 
As in hydrodynamic turbulence, we drive turbulence in two wavenumber ranges, $k_L\sim 2$ and $k_S\sim 20$, in
Fourier space.

\subsection{Turbulence with a Weak Mean Magnetic Field}  \label{sect:weakB}
In this subsection, we consider driven turbulence permeated by a weak mean magnetic field,
the strength of which is set to 0.001.

\subsubsection{Simulations with $\epsilon_L/\epsilon_S < 1$}   
\label{sect:sd}

We perform simulations of incompressible MHD turbulence with identical 
small-scale driving and different large-scale driving. 
Therefore the small-scale energy-injection rate $\epsilon_S$ is similar for all runs,
but the large-scale energy-injection rate $\epsilon_L$ varies.
The simulation parameters  are
listed in Table 1. 

\begin{figure*}
\begin{tabbing}
 \includegraphics[width=0.40\textwidth]{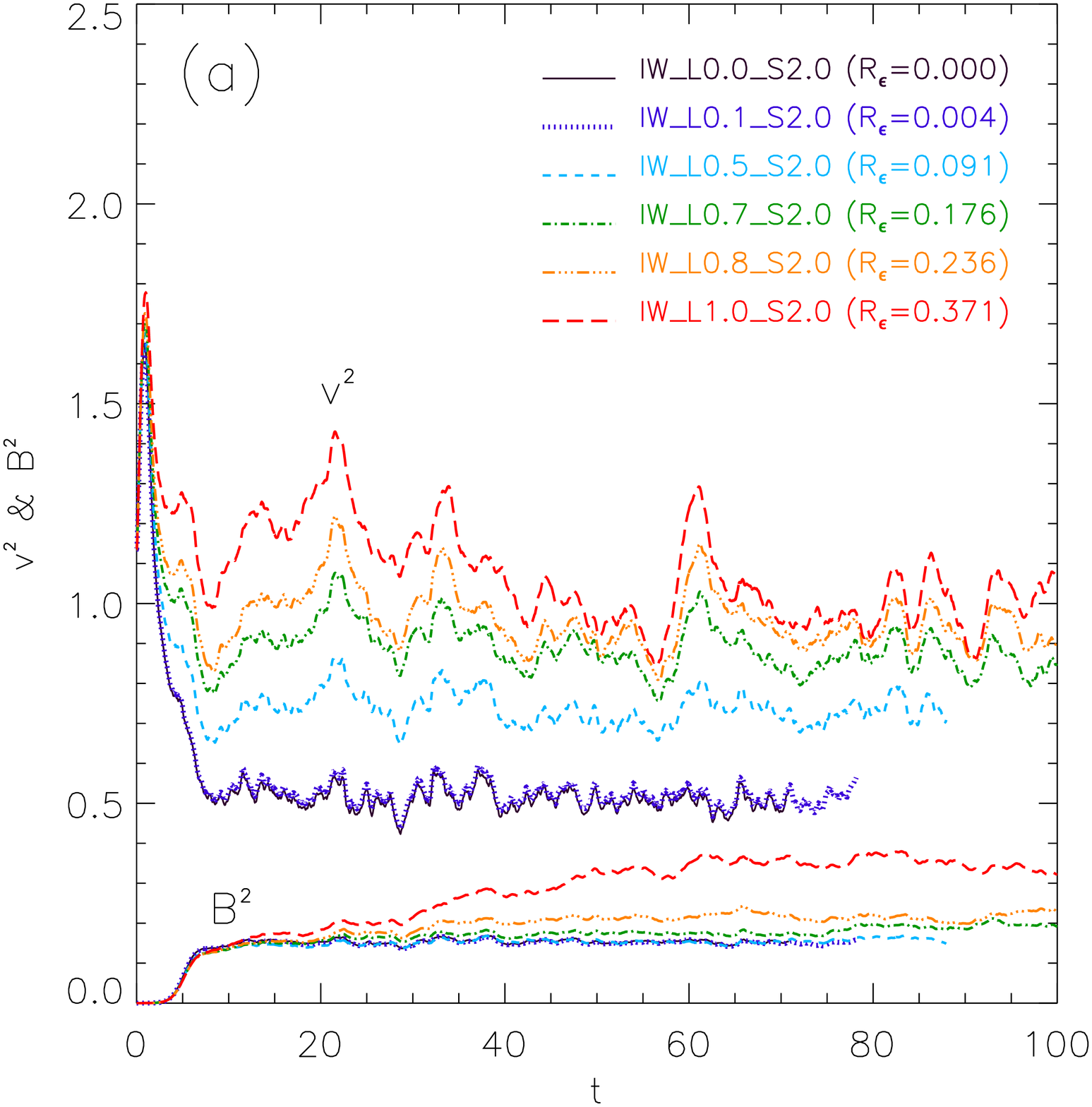}
 \=
  \includegraphics[width=0.40\textwidth]{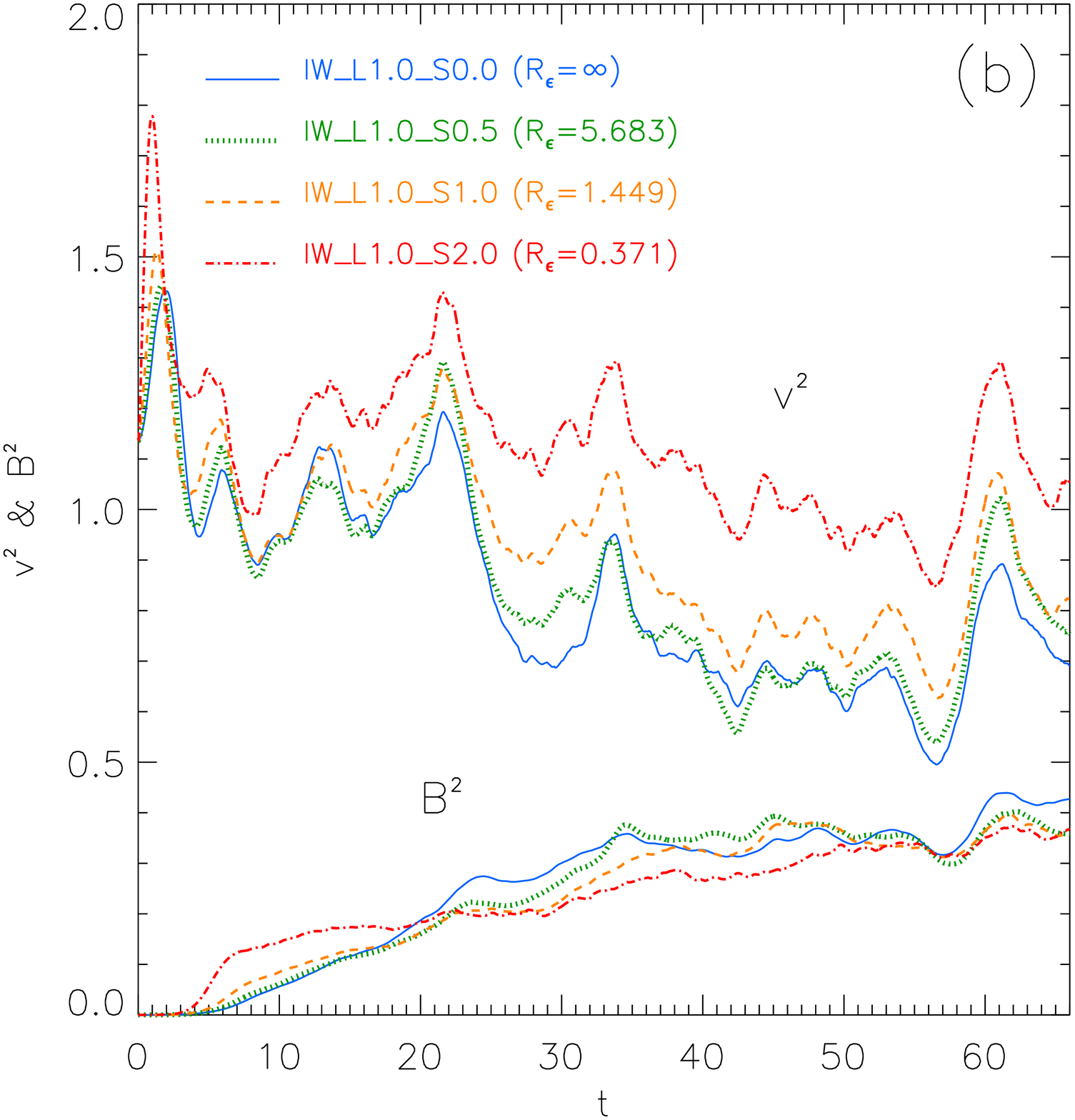} \\
  \includegraphics[width=0.40\textwidth]{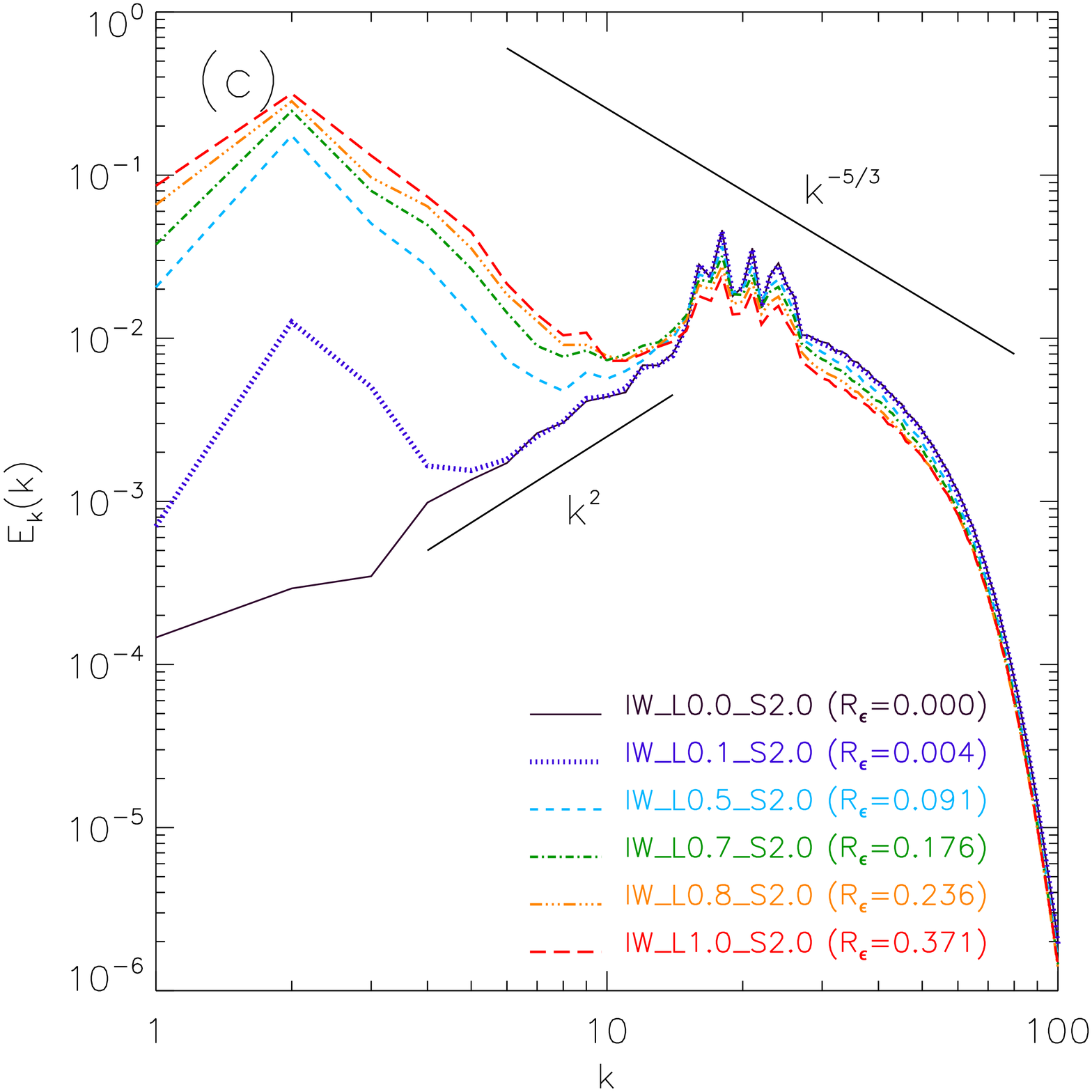}
  \>
  \includegraphics[width=0.40\textwidth]{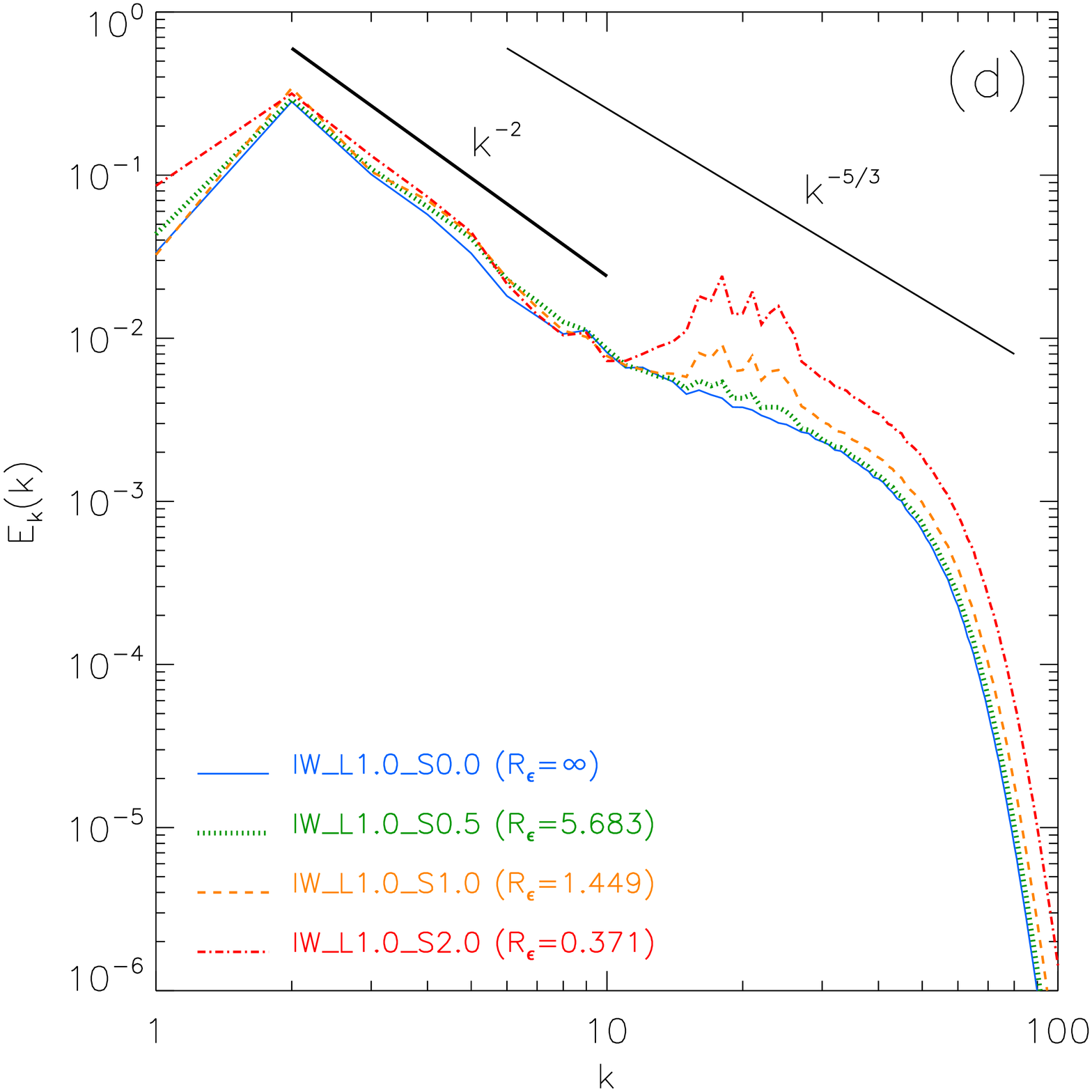} \\
  \includegraphics[width=0.40\textwidth]{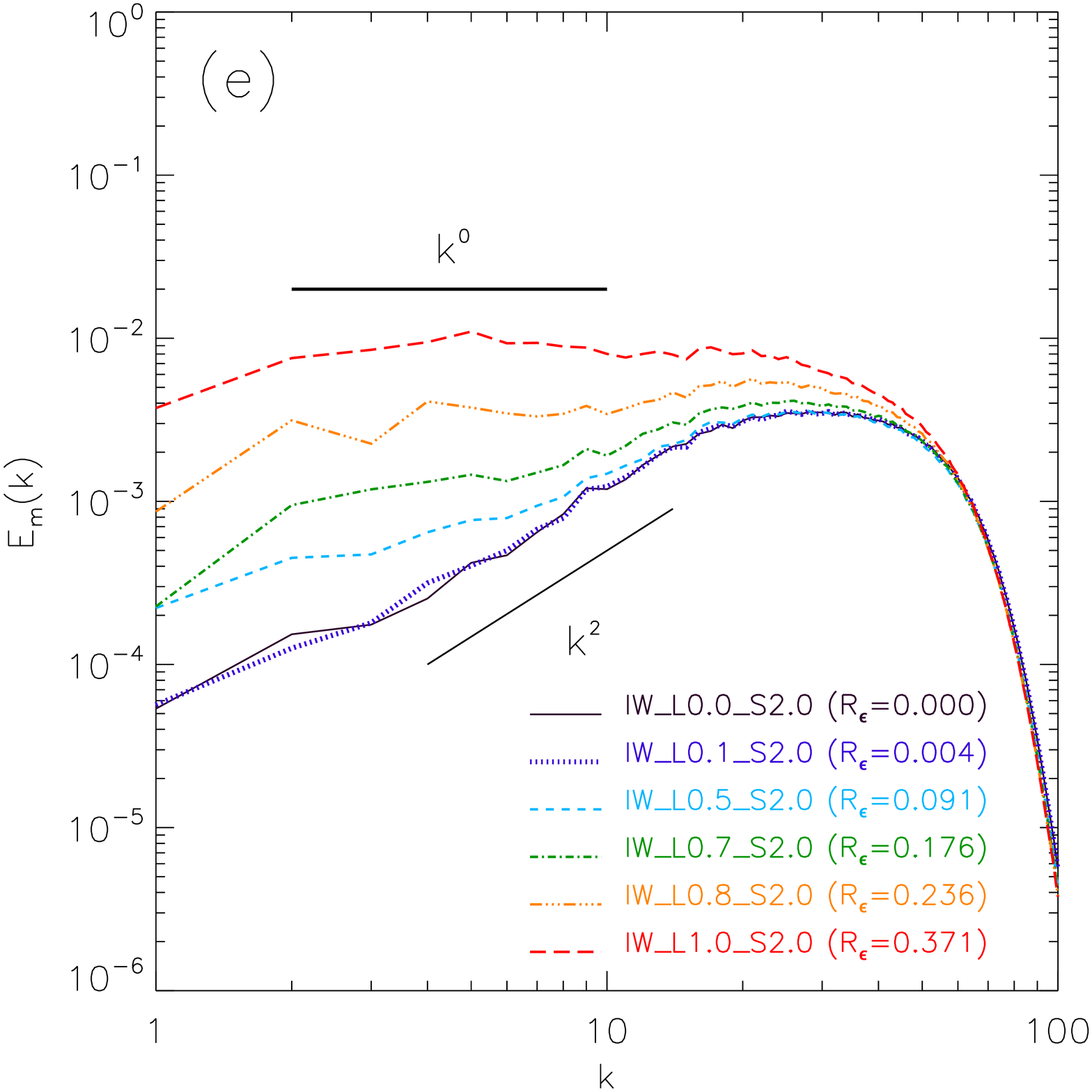}
  \>
  \includegraphics[width=0.40\textwidth]{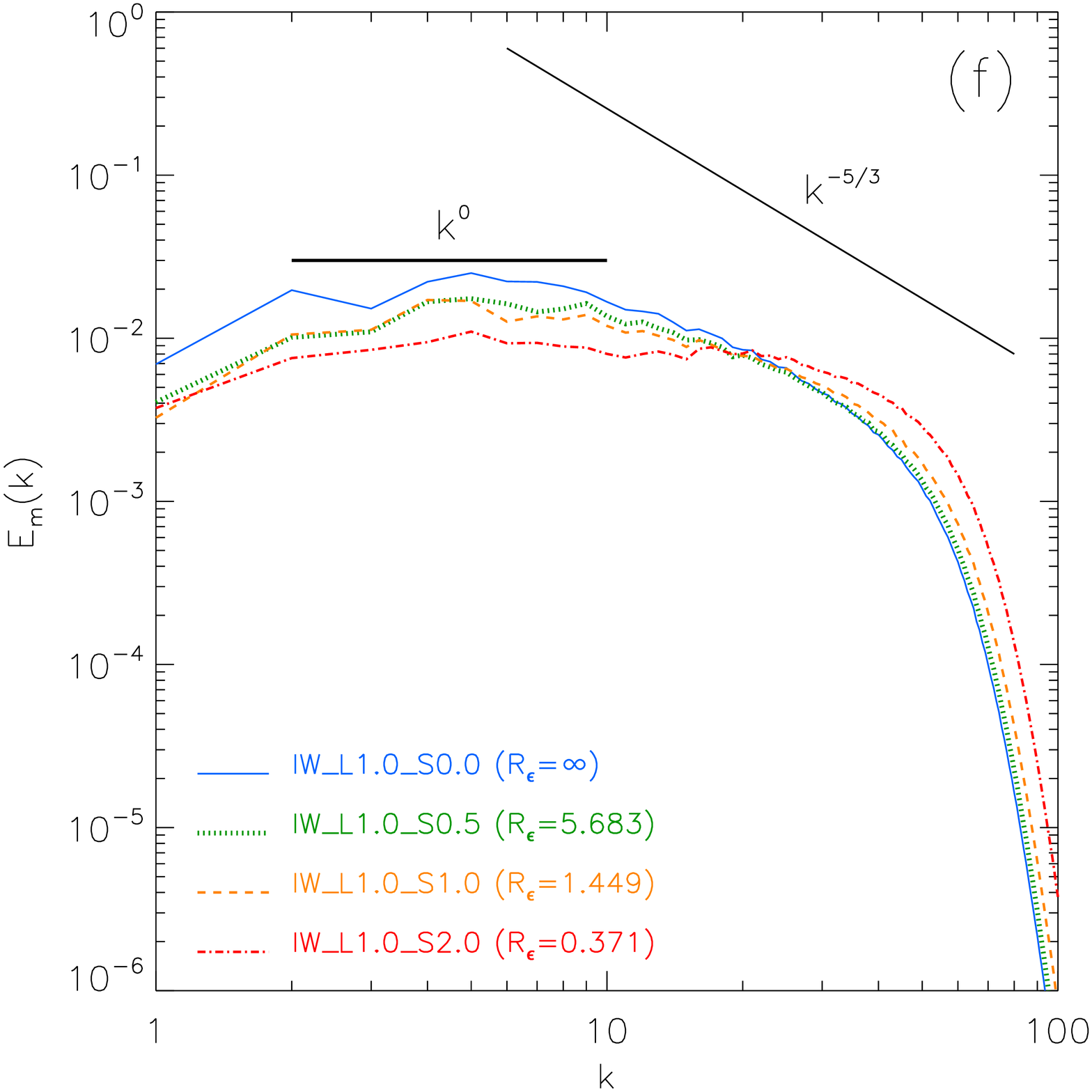} 
  \end{tabbing}
  \caption{
  Incompressible MHD turbulence simulations with a weak mean magnetic field.
  Panels in the left column are for $\epsilon_L/\epsilon_S < 1$ and
  those in the right column are for $\epsilon_L/\epsilon_S \gtrsim 1$.
Top panels: time evolution of kinetic and magnetic energy densities. 
Middle panels: kinetic energy spectra at t$\sim$ 66.
Bottom panels: magnetic energy spectra at t$\sim$ 66.
   }
\label{f:weakB}
\end{figure*}

Figure~\ref{f:weakB}(a) shows kinetic  (upper curves) and magnetic (lower curves) energy   densities 
as functions of time. 
The purple solid lines correspond to Run $IW\_L0.0\_S2.0$, in which turbulence is driven only at 
the small scale ($k_S \sim$20).
In other runs, turbulence is driven at both the large scale ($k_L\sim$2) and the small scale ($k_S \sim$20).

The upper curves show the time evolution of the kinetic energy densities. 
The lowest curve among the upper curves corresponds to
the run with only the small-scale driving ($IW\_L0.0\_S2.0$; the solid line).
When $R_\epsilon$ is very small (see the blue dotted line; $IW\_L0.1\_S2.0$) the time evolution
of kinetic energy density is almost indistinguishable from that of the single-driving case.
Indeed the blue dotted line ($R_\epsilon \sim 0.004$) and the purple solid line ($R_\epsilon \sim 0$)
almost coincide.
For larger values of $R_\epsilon$, the kinetic energy density clearly goes up when 
we inject more energy into the large scale. 
In Run $IW\_L0.5\_S2.0$ (the cyan dashed line), the large-scale energy-injection rate is 
only $\sim$9\% of the small-scale one.
Nevertheless, the kinetic energy density of Run $IW\_L0.5\_S2.0$
is much larger than that of  the single-driving case 
(Run $IW\_L0.0\_S2.0$; the purple solid line).
This is because $k_S/k_L$ is large.   
In fact, this result is consistent with Equation~(\ref{eq:vv}), 
which states that $v_L/ v_S \sim (k_S/k_L)^{1/3} (\epsilon_L/\epsilon_S)^{1/3} \sim 2.15 (\epsilon_L/\epsilon_S)^{1/3}$ in our simulations. 

The behavior of the magnetic energy density is different from that of kinetic energy density.
It seems that, if $R_\epsilon$ is less than a certain value, which is $\sim 0.1$
in our case, 
the additional large-scale driving
does not influence the magnetic energy density much.
The lower curves in Figure \ref{f:weakB}(a) show the time evolution of the magnetic energy densities.
Let us compare three runs, ``$IW\_L0.0\_S2.0$'', ``$IW\_L0.1\_S2.0$'' and ``$IW\_L0.5\_S2.0$'',
in which the large-scale energy-injection rates are less than $\sim$9\% of the small-scale ones.
The time evolutions of magnetic energy densities of 
 the three runs, the  lowest three curves in Figure \ref{f:weakB}(a),
almost coincide.
They initially rise quickly as a result of the turbulence dynamo effect 
by the strong fluid motions near $l_S$.
Then, after $t\sim 7$, they become saturated.
All three runs show
almost identical behaviors.
In runs with stronger large-scale driving, magnetic energy densities 
gradually reach higher 
levels of saturation due to the additional turbulence dynamo effect by the motions near $l_L$.
The saturation level for the magnetic energy density is less sensitive to $R_\epsilon$ compared with
that for kinetic energy density.
This might imply that enhanced turbulence diffusion by small-scale driving prevents the
magnetic energy density from growing as fast as the kinetic energy density.
We will discuss the turbulence dynamo in the presence of two-scale driving in Section \ref{sect:turb_dyn}.

We plot kinetic energy spectra in Figure~\ref{f:weakB}(c) (middle-left panel). 
Since we inject most of the energy at $k_S\sim$20, we see a spectral peak at $k\sim k_S$ in all runs.
When there is energy injection only at $k_S \sim$20 (Run $IW$\_$L0.0$\_$S2.0$), 
the kinetic energy spectrum for k$<$20 is compatible with $\sim k^2$.
When there is an additional energy injection at $k_L \sim$2, we can see an additional spectral peak
at $k\sim k_L$.
In the plot, we can clearly see a peak at $k\sim k_L$ even though 
the large-scale energy-injection rate is very small.
For example, in case of Run $IW$\_$L0.1$\_$S2.0$ (the blue dotted line), $\epsilon_{L}$ 
is smaller than $\epsilon_{S}$ by three orders of magnitude (see also Table 1). 
Nevertheless, we can clearly see a peak at $k_L\sim$2.
As we inject more energy into the large scale, 
the spectral peak at the large scale goes up. 
The kinetic spectrum 
between $k_L$ and $k_S$
is somewhat steeper than the Kolmogorov one, which might stem from two effects. 
First, when the external magnetic field is weak it is common to
observe a steep large-scale kinetic spectrum (see, for example, Cho et al.~2009).
Second, a fast energy cascade caused by the strong turbulence motions at $k_S\sim$20 can also make
the large-scale  kinetic spectrum steeper.

In Figure~\ref{f:weakB}(e) (lower-left panel), we compare magnetic energy spectra. 
Unlike the kinetic energy spectra, the magnetic spectra do  not clearly show double peaks\footnote{
  This result may change if the two driving scales are well separated. 
  In our simulations, $k_S/k_L \sim 10$.
  If $k_S/k_L \gg 10$, it may be possible to see two peaks in the magnetic spectrum. 
}.
When there is energy injection only at $k_S\sim$20 (Run $IW$\_$L0.0$\_$S2.0$; $R_\epsilon=0$), 
the magnetic energy spectrum shows a peak at a wavenumber a bit larger than $k_S$ and
the spectrum for k$<$20 is compatible with $\sim k^2$.
As additional energy is injected
into the large scale, the large-scale magnetic energy increases and, as a result, 
the magnetic energy spectrum between $k_L\sim$2 and $k_S\sim$20  becomes flatter.
In the case of Run $IW$\_$L1.0$\_$S2.0$, in which $R_\epsilon \sim$0.371, 
the magnetic spectrum between the two wave-numbers is nearly flat.

\begin{figure}[t]
\centerline{\includegraphics[width=.5\textwidth]{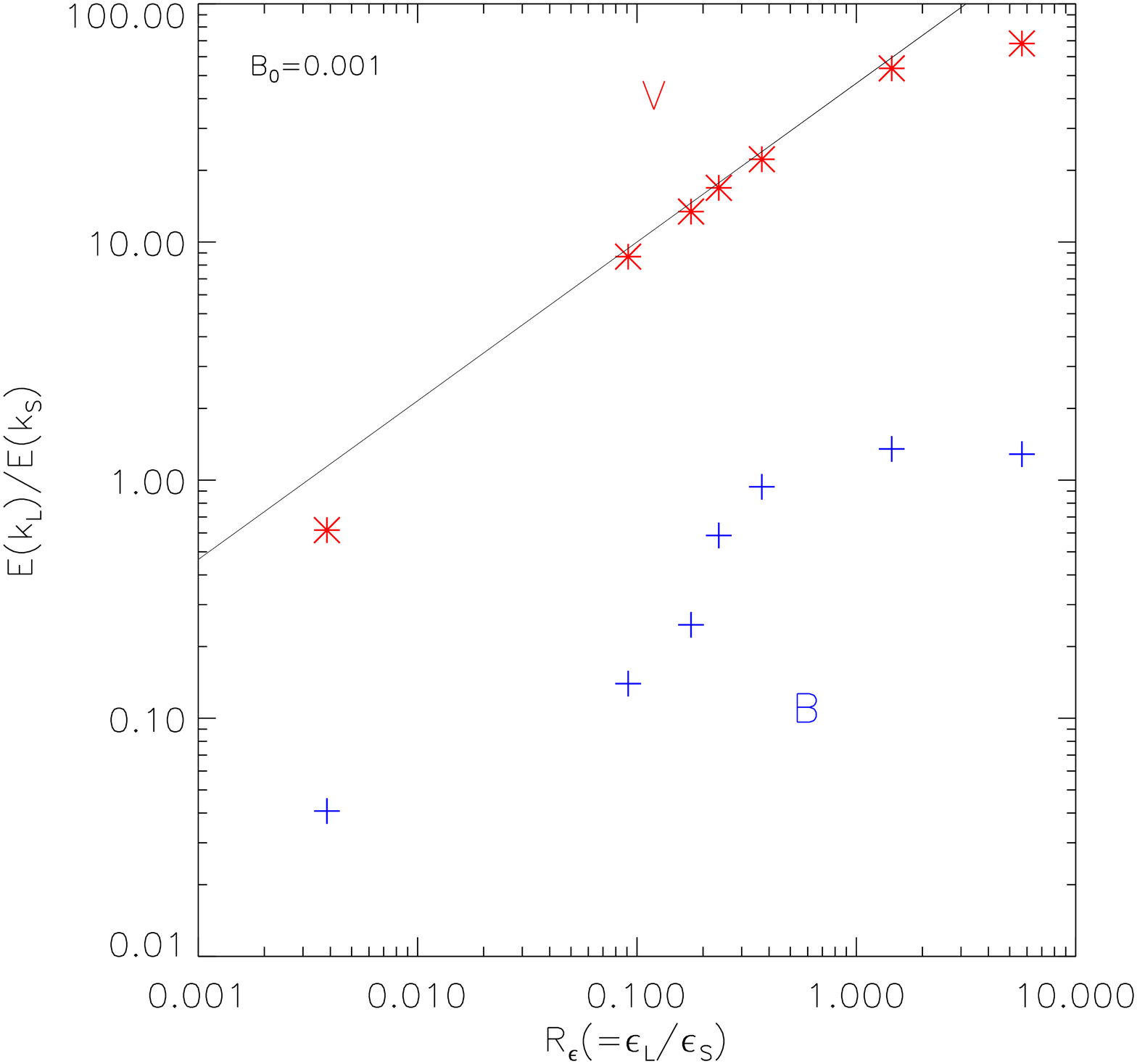}}
\caption{
  Relation between $E(k_L)/E(k_S)$ and $\epsilon_L/\epsilon_S$ for
  runs with a weak mean magnetic field.
The red asterisks are for the velocity field and
the blue plus symbols are for magnetic field. 
The black solid line indicates Eq. (\ref{eq:theo_exp}).
Results are for t$\sim$66.}
\label{f:pbp-wksf}
\end{figure}

In Section \ref{sect:3.2} (see Equation (\ref{eq:theo_exp}); see also Equation~(\ref{eq:EE})), 
we derived 
the relation between the ratio of energy-injection rates $R_\epsilon$ (= $\epsilon_L/\epsilon_S$)
and the ratio of energy spectra at the peaks $E(k_L)/E(k_S)$. 
For the former, we directly calculate ${\bf f}\cdot {\bf v}$ at two driving scales (see Table 1).
For the latter, we take $k_L=$2 and $k_S=$20. 
We present the relation between two ratios in Figure~\ref{f:pbp-wksf}.  
The asterisks indicate the velocity field and the plus symbols indicate the magnetic field.
The solid guide line is the expected relation in Eq. (\ref{eq:theo_exp}).
The results for the velocity field  follow the guide line well.

However, results for the magnetic field do not follow the guide line. 
For a given $\epsilon_L/\epsilon_S$,
$E(k_L)/E(k_S)$ for the magnetic field is an order of magnitude smaller than that for the velocity,
which is mainly due to the fact that the increase in the magnetic spectrum at $k_L$ is notably smaller than
that of kinetic spectrum (compare Figures~2(c) and (e)).
The magnetic spectrum at $k_L$  increases relatively slowly as we increase $\epsilon_L/\epsilon_S$ because
turbulence diffusion by motions near $l_S$ 
prevents large-scale magnetic fields from growing.
The fact that the magnetic spectrum near $k_S$   
 increases as $\epsilon_L/\epsilon_S$ increases also contributes to
the smallness of $E(k_L)/E(k_S)$.
The magnetic spectrum at $k_S$ increases as we increase $\epsilon_L/\epsilon_S$ because
fluctuating magnetic field near $l_S$ increases as the \textit{local} mean magnetic field
increases, which is caused by enhanced $E(k_L)$.

\subsubsection{Simulations with $\epsilon_L/\epsilon_S \gtrsim 1$}  \label{sect:ld}
In the previous sub-subsection, we considered the cases of $\epsilon_L/\epsilon_S <1$ and studied
how the properties of turbulence changed as
we increased the value of $\epsilon_L/\epsilon_S$.
For that purpose,
we fixed the amplitude of small-scale driving
 in all runs and
considered the effects of large-scale driving as we increased the strength of the
large-scale driving. 
In this sub-subsection, we investigate the properties of turbulence for $\epsilon_L/\epsilon_S \gtrsim 1$.
For the sake of numerical convenience, 
we fix the amplitude of large-scale driving in all simulations and
use different strengths of small-scale driving. 
Therefore, in this sub-subsection, we first consider the case of single driving at $k_L\sim2$.
Then, we consider the effects of additional small-scale driving at $k_S\sim 20$.
The resulting $\epsilon_L/\epsilon_S$ ratio ($= R_\epsilon$) is listed in Table 2.

Figure \ref{f:weakB}(b) (upper-right panel) shows time evolution of kinetic (upper curves) 
and magnetic (lower curves) 
energy densities. 
Initially, $v^2 \gtrsim 1$ and $B^2 = B_0^2=10^{-6}$.
As time goes on, $B^2$ grows and $v^2$ drops gradually.
When the large-scale driving dominates (i.e.~$R_\epsilon >1$), $v^2$ is not much 
different from that of the single-driving case (Run $IW\_L1.0\_S0.0$; $R_\epsilon =\infty$).
However, when the small-scale driving dominates, 
so that $R_\epsilon <1$ (Run $IW\_L1.0\_S2.0$; the red dot-dashed line),
$v^2$ becomes noticeably larger.
In general, $v^2$ becomes larger when the small-scale forcing gets stronger.
However, the saturation levels of $B^2$ do not show strong dependence on the strength of 
small-scale driving.

The lower curves in Figure~\ref{f:weakB}(b) show that, 
when the large-scale driving dominates (Runs $IW\_L1.0\_S0.5$ and $IW\_L1.0\_S1.0$;~$R_\epsilon >1$), 
the time evolution of the magnetic energy density is very similar to that of 
 the run with only the large-scale driving (Run $IW\_L1.0\_S0.0$; the blue solid line).
However, when the small-scale driving becomes dominant (Run  $IW\_L1.0\_S2.0$; $R_\epsilon <1$), 
we see a fast increase in the magnetic energy density for $t \lesssim 6$.
This is due to the fast stretching of the magnetic field lines by the small-scale motions, which
gets saturated when the magnetic energy density becomes comparable to kinetic
energy density at the small driving-scale.
After saturation at the small-scale, it seems that an additional growth stage caused by the large-scale
driving follows. 
It is interesting that the growth rate during the additional growth stage is smaller
than that of the single-driving case
(compare Run $IW\_L1.0\_S2.0$ and Run $IW\_L1.0\_S0.0$).
We believe that this is due to enhanced turbulence diffusion caused by the strong turbulence motions
at the small driving scale.
Owing to the smaller linear growth rate, the system reaches a final saturation stage later.
Figure~\ref{f:weakB}(b) shows the magnetic energy densities are all similar
during the saturation stage in all runs, 
which implies that additional small-scale driving does not significantly affect 
the final states of turbulence.

In Figure~\ref{f:weakB}(d) (middle-right panel), 
we plot the kinetic energy spectra during the  saturation stage. 
The slopes of kinetic energy spectra for 2$\leq$k$\leq$10
 are nearly the same in all runs and are steeper than Kolmogorov's~-5/3 slope. 
As we discussed in the earlier sub-subsection,
a steep kinetic energy spectrum near the driving scale 
is a commonly observed feature in ``super-Alfvenic" turbulence, 
MHD turbulence with a weak mean magnetic field, driven at a single scale.
(e.g.~Cho et al. 2009).

The peaks of the kinetic energy spectra at $k_L\sim$2 almost coincide because we
use the same large-scale forcing in those runs.  
The peak at $k_S\sim$ 20 may or may not appear depending on the energy-injection rate
at the small scale.
In order to have a visible peak at the small scale,
energy-injection rate at the small scale should be larger than that at the large scale 
(see Eq.~(\ref{eq:9})). 
Actually, the small-scale peak in the kinetic spectrum at $k_S\sim$20 can be visible  even though 
the small-scale energy-injection rate is slightly smaller than the large-scale one
(Run $IW$\_$L1.0$\_$S1.0$; $R_\epsilon \sim 1.449$).
For a smaller $R_\epsilon$,
it may be difficult to say whether or not we have two peaks in the kinetic energy spectrum.

Figure~\ref{f:weakB}(f) (lower-right panel) shows comparison of the magnetic energy spectra after saturation. 
When we have only the large-scale driving (see the blue solid line), 
the magnetic spectrum has a peak at a wavenumber about 2 to 3 times 
larger than the driving-scale wavenumber, 
which is consistent with earlier simulations (see, for example, Cho \& Vishniac 2000; Cho et al. 2009).
Even though we apply additional driving at $k_L\sim$20, the shape of magnetic spectra 
does not change much if
the small-scale energy-injection rate is smaller than the large-scale one, i.e.~$R_\epsilon >1$.
However, if the small-scale energy-injection rate is larger than the large-scale one 
(Run  $IW\_L1.0\_S2.0$; $R_\epsilon <1$),
we can see that the magnetic spectrum near $k_L$ goes down and that the overall spectrum
becomes shallower than that of the single-driving case.\footnote{
    If we inject more energy at the small scale, the slope will increase and become positive (see
   \S\ref{sect:sd}).
}

\subsection{Turbulence with a Strong Mean Magnetic Field}  
\label{sect:strongB0}

In this section, we consider strongly magnetized incompressible turbulent fluids. 
The strength of the mean magnetic field $B_0$ is set to 1.0.
Since we are mainly interested in the behavior of velocity and magnetic field on large scales,
we use identical small-scale driving in all runs in this subsection.  
As we list in Table 3, 
small-scale driving dominates large-scale driving in all runs. 

\begin{figure*}
\begin{tabbing}
 \includegraphics[width=0.42\textwidth]{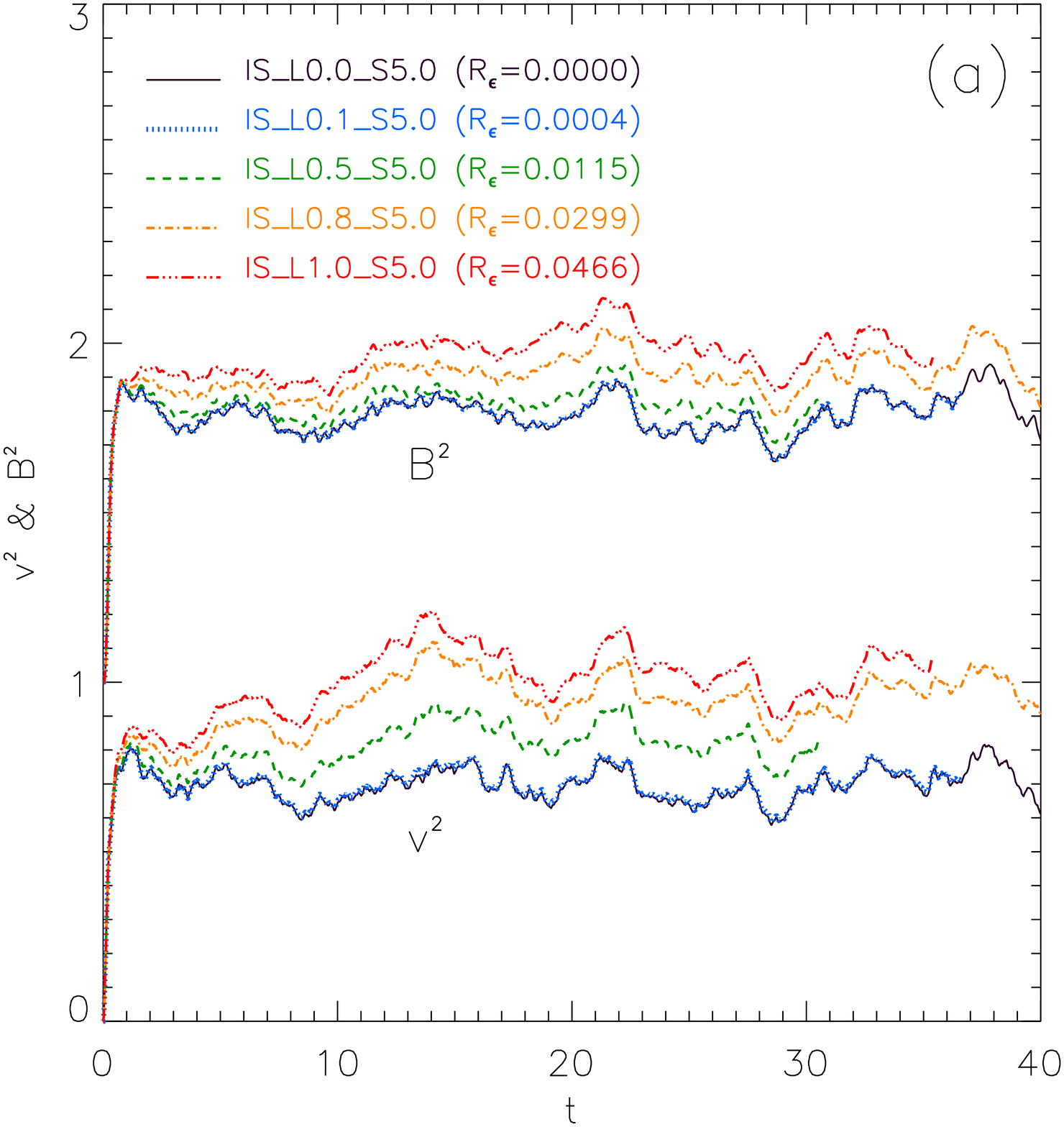}
 \=
  \includegraphics[width=0.42\textwidth]{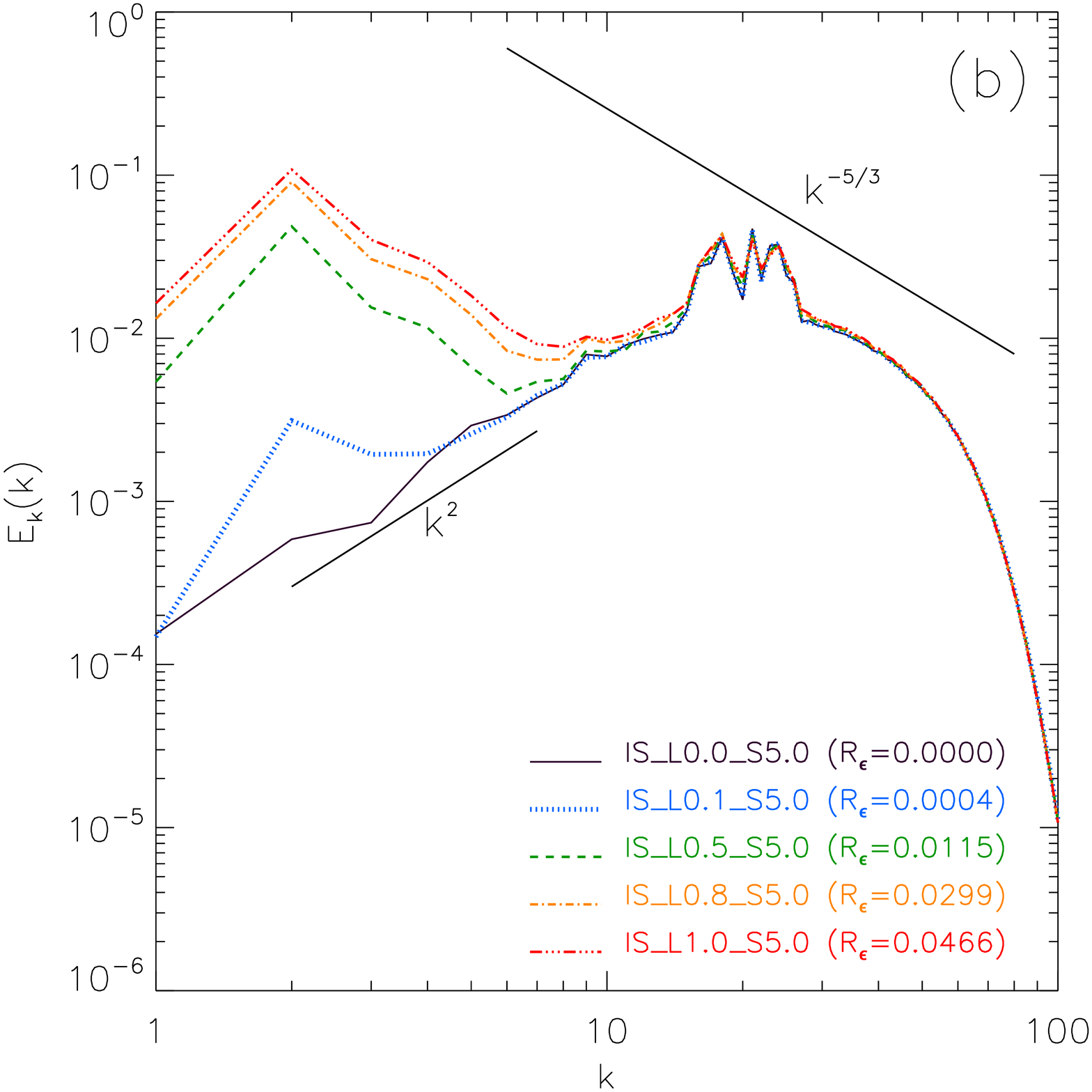} \\
  \includegraphics[width=0.42\textwidth]{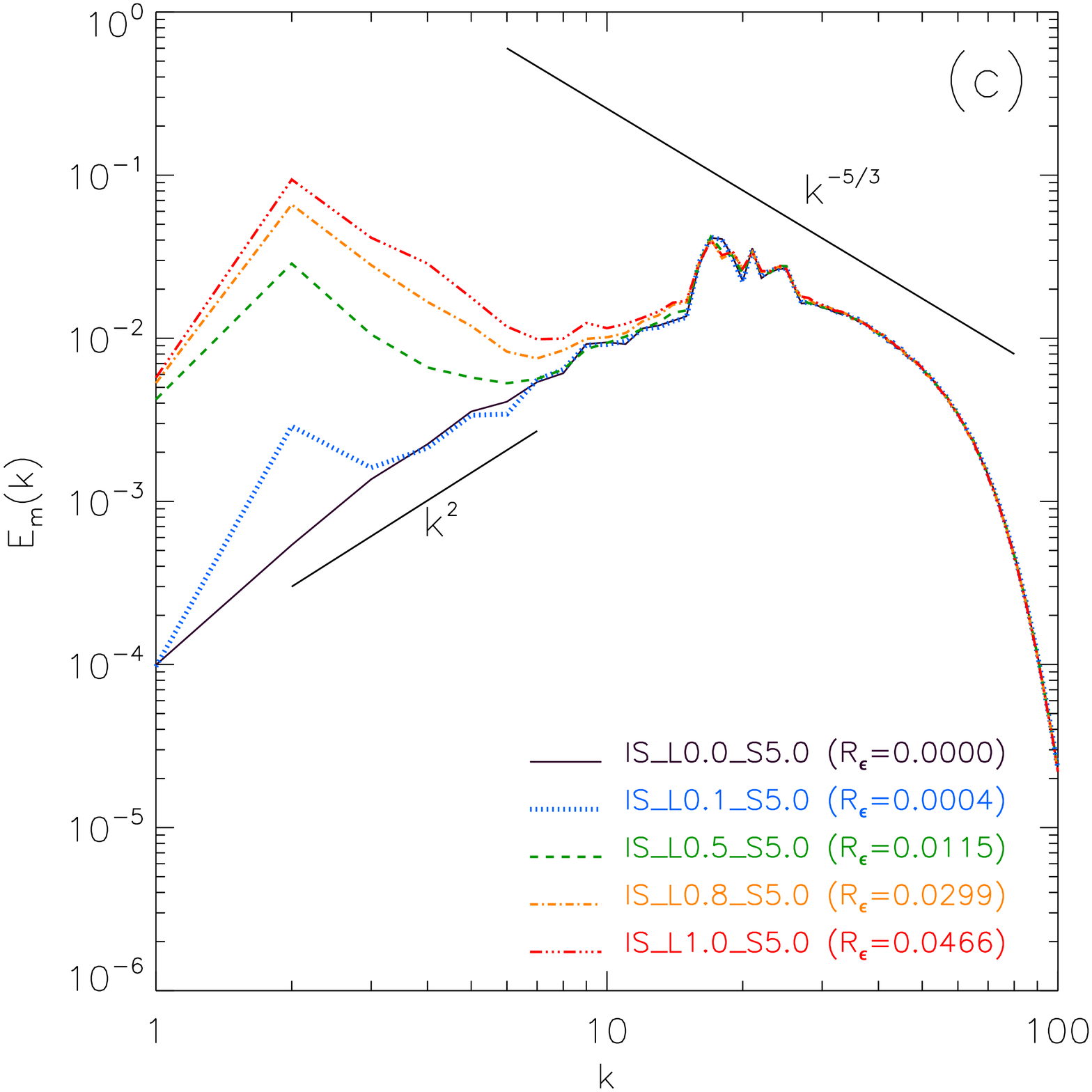}
  \>
  \includegraphics[width=0.42\textwidth]{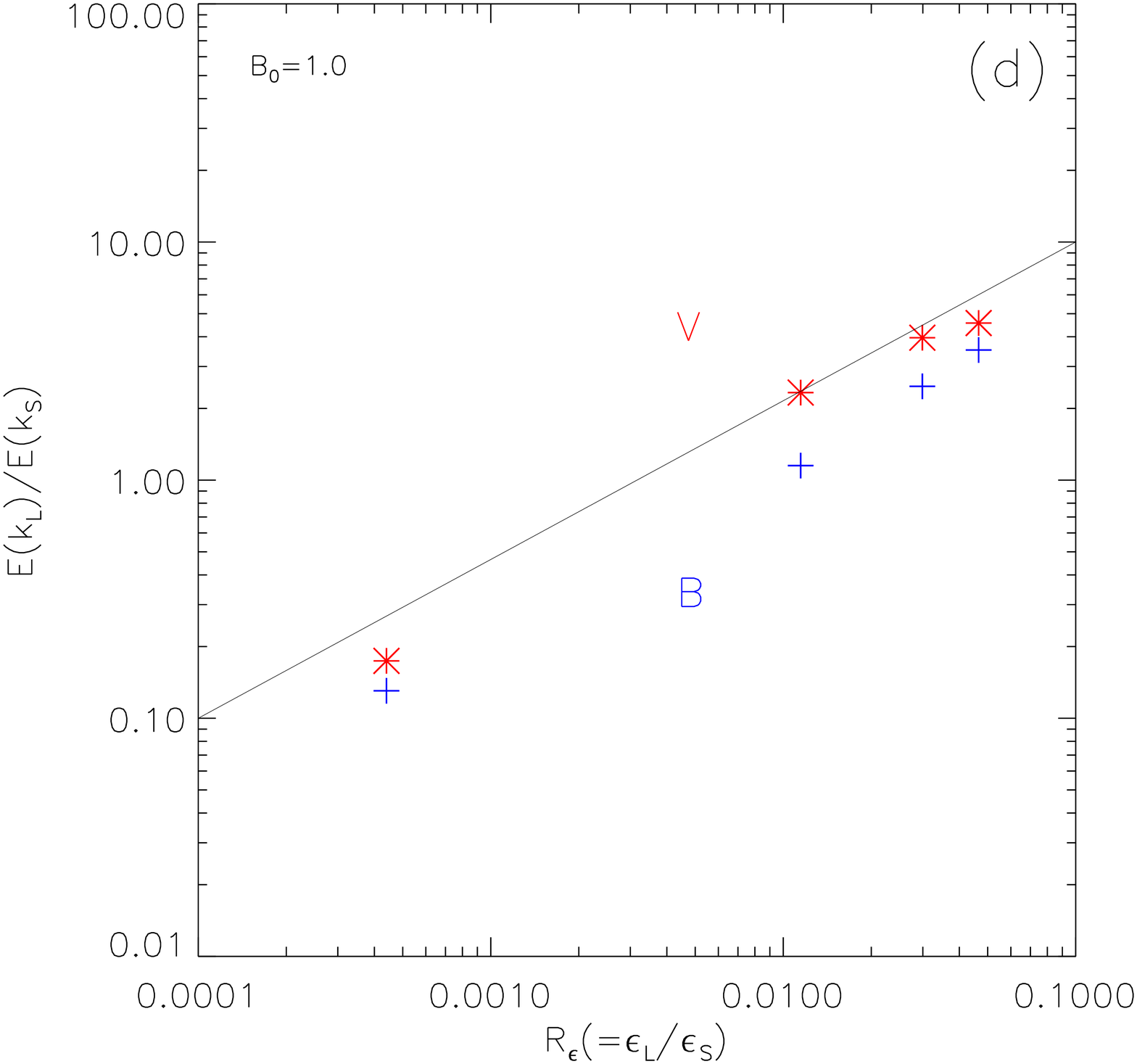} 
  \end{tabbing}
  \caption{Incompressible MHD turbulence simulations with a strong mean magnetic field.
  (a) Time evolution of kinetic and magnetic energy densities. 
  (b) Kinetic energy spectra at t$\sim$ 31. 
  (c) Magnetic energy spectra at t$\sim$ 31. 
  (d) The relation between $E(k_L)/E(k_S)$ and $\epsilon_L/\epsilon_S$.
The red asterisks are for the velocity field and the blue plus symbols are for the magnetic field. 
The black solid line indicates Eq. (\ref{eq:theo_exp}).
Results are for t$\sim$31.  
}
\label{f:strongB}
\end{figure*}

In Figure~\ref{f:strongB}(a), 
we plot the time evolution of $v^2$ (lower curves) and $B^2$ (=$B_0^2 +b^2$; upper curves) for five different models. 
At t=0, ${\bf v}={\bf b}=0$ in all simulations.
As the simulations start, both the kinetic and magnetic energy densities grow quickly.
Due to the strong external magnetic field, 
the fluids reach saturation quickly.
The purple solid lines are for the run with a single driving at $k_S\sim$20.
When we inject additional energy at $k_L\sim$2, both the kinetic and magnetic energy densities
increase. 
Note that, even though we inject a very small fraction of energy into the large scale,
the kinetic and magnetic energy densities go up substantially (see, for example, Run $IS\_L0.5\_S5.0$;
the green dashed lines).

We plot the kinetic and magnetic energy spectra 
in Figures~\ref{f:strongB}(b) and (c), respectively. 
 In both the kinetic and magnetic energy spectra, 
we can observe two peaks, one at $\sim k_L$ and the other at $\sim k_S$. 
Since we use the same small-scale driving in all runs, the heights of the spectral peak at $k_S\sim$20
are all similar.
However, since we use different strengths of the large-scale forcing, the heights of the spectral peaks at $k_L\sim$2
are different. 
As in  the weak mean-field cases (Section \ref{sect:weakB}), we can observe the large-scale spectral peak
even in the case that the large-scale energy-injection rate is a small fraction of the small-scale one.

The velocity and magnetic field show very similar spectra, which
is due to the strong mean magnetic field. 
When the medium is permeated by a strong mean magnetic field, fluctuations act like 
 Alfv\'en waves, in which magnetic fluctuations are coupled with velocity fluctuations.
Spectral slopes in the inertial range are close to Kolmogorov's $-5/3$ slope. 

As in the previous subsection, we plot in  Figure~\ref{f:strongB}(d)
the relation between the ratio of energy-injection rates $\epsilon_L/\epsilon_S$
and the ratio of energy spectra at the peaks $E(k_L)/E(k_S)$. 
 In case of strongly magnetized turbulence, 
not only the kinetic energy spectrum
but also the magnetic energy spectrum show fairly good agreement with the theoretical expectation in Eq.~(\ref{eq:theo_exp}).

\section{Density Fluctuations in Compressible MHD Turbulence}

\subsection{Density Fluctuations}
Density fluctuations are of great importance in many astrophysical problems, such as star formation and interstellar radio scintillations.
It is well known that Mach number affects the slope of the density spectrum 
(Beresnyak, Lazarian,\&  Cho 2005; Kim \& Ryu 2005).
Kim $\&$ Ryu (2005) showed how  the slope of density power spectrum changes as
the r.m.s.\ Mach number increases in compressible hydrodynamic turbulence
with a single energy-injection scale.

\begin{figure*}
\begin{tabbing}
 \includegraphics[width=0.40\textwidth]{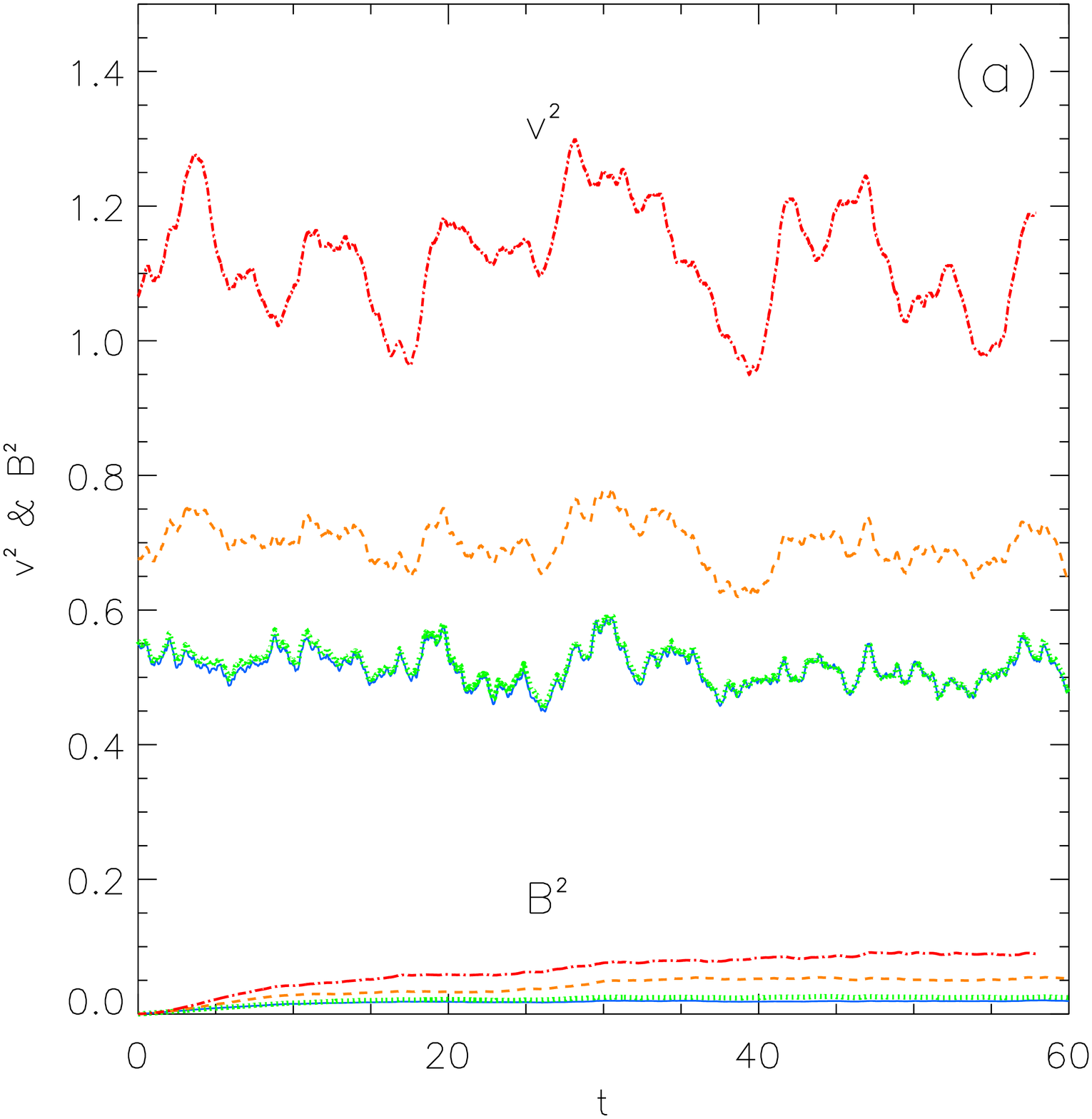}
 \=
  \includegraphics[width=0.40\textwidth]{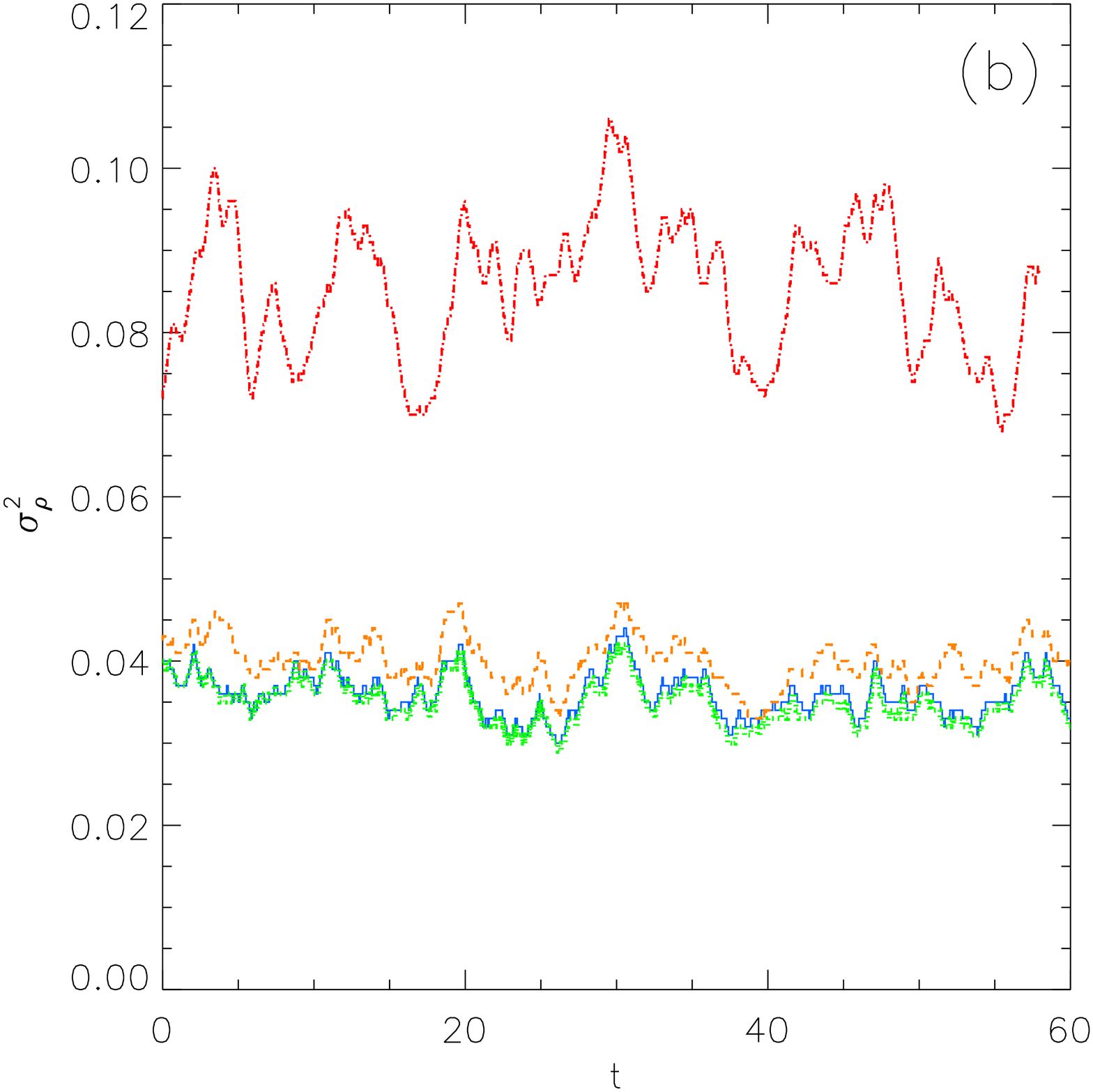} \\
  \includegraphics[width=0.40\textwidth]{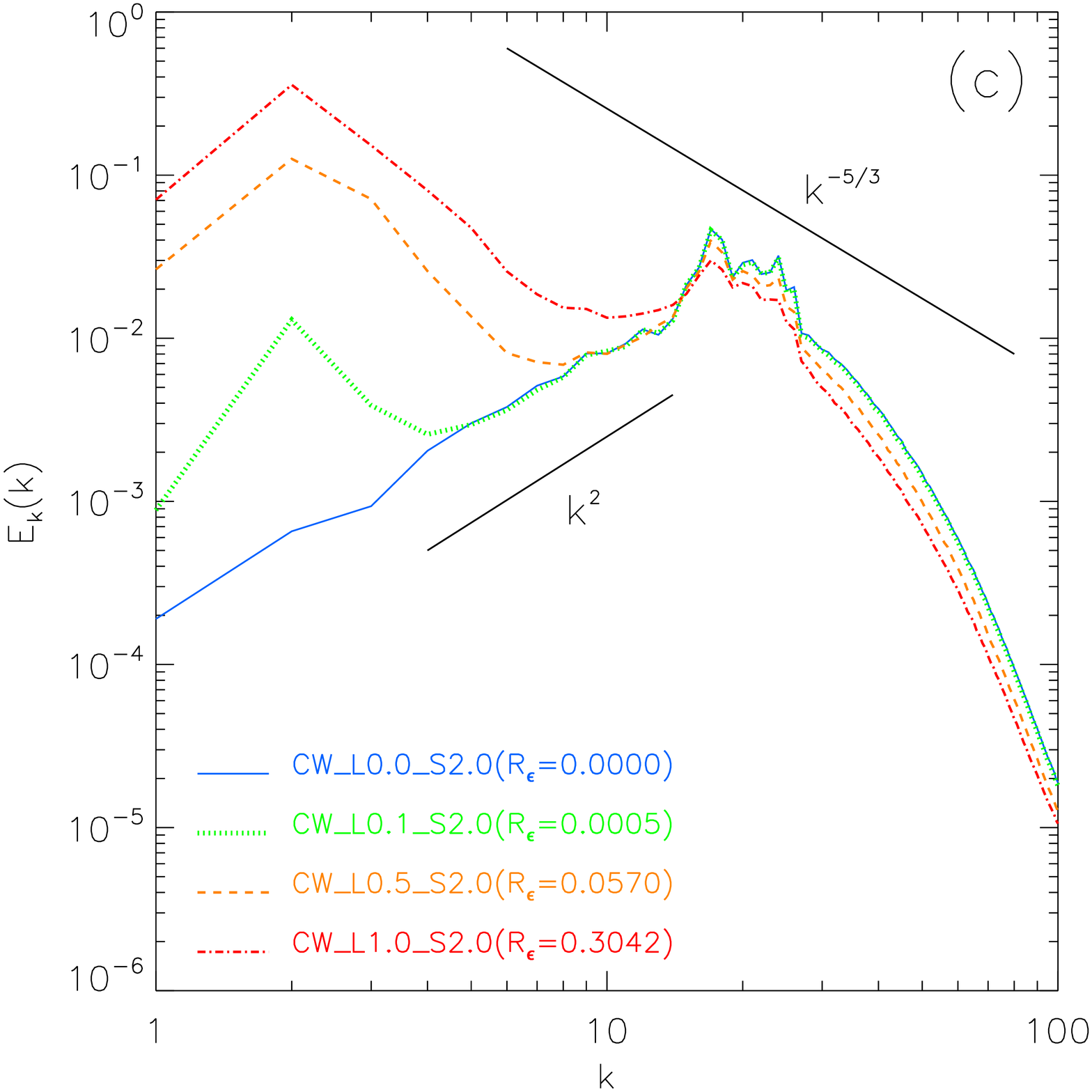}
  \>
  \includegraphics[width=0.40\textwidth]{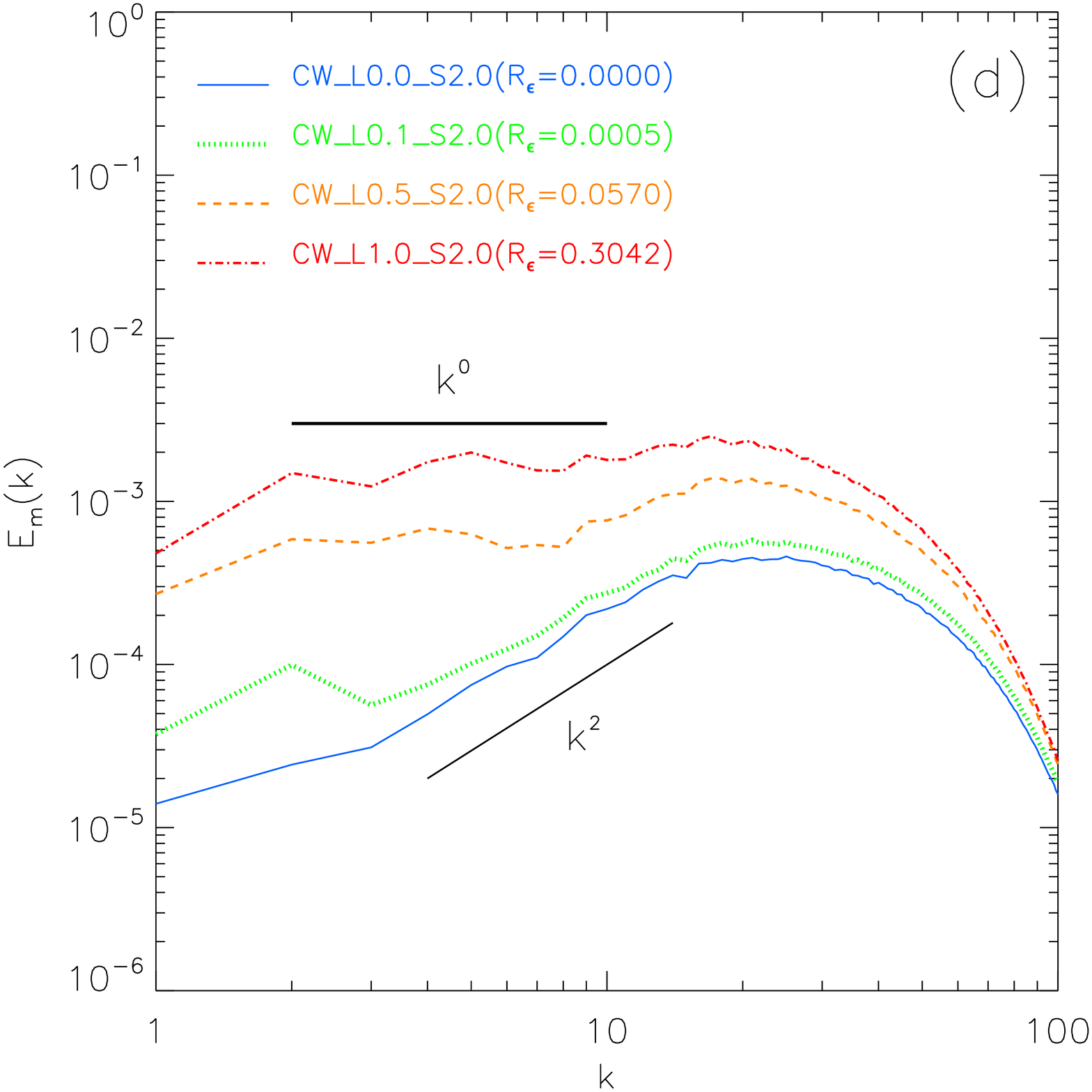} \\
  \includegraphics[width=0.40\textwidth]{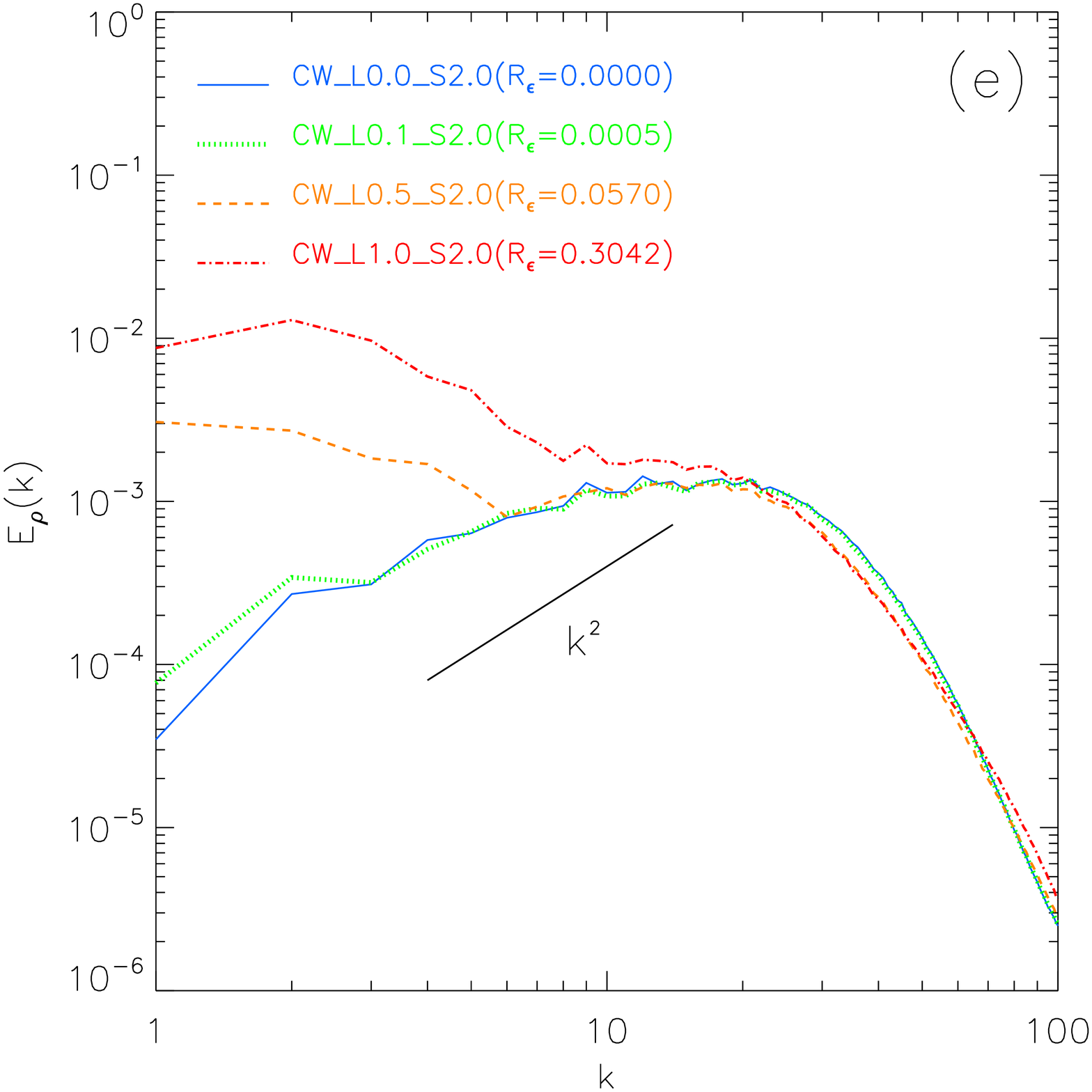}
  \>
  \includegraphics[width=0.40\textwidth]{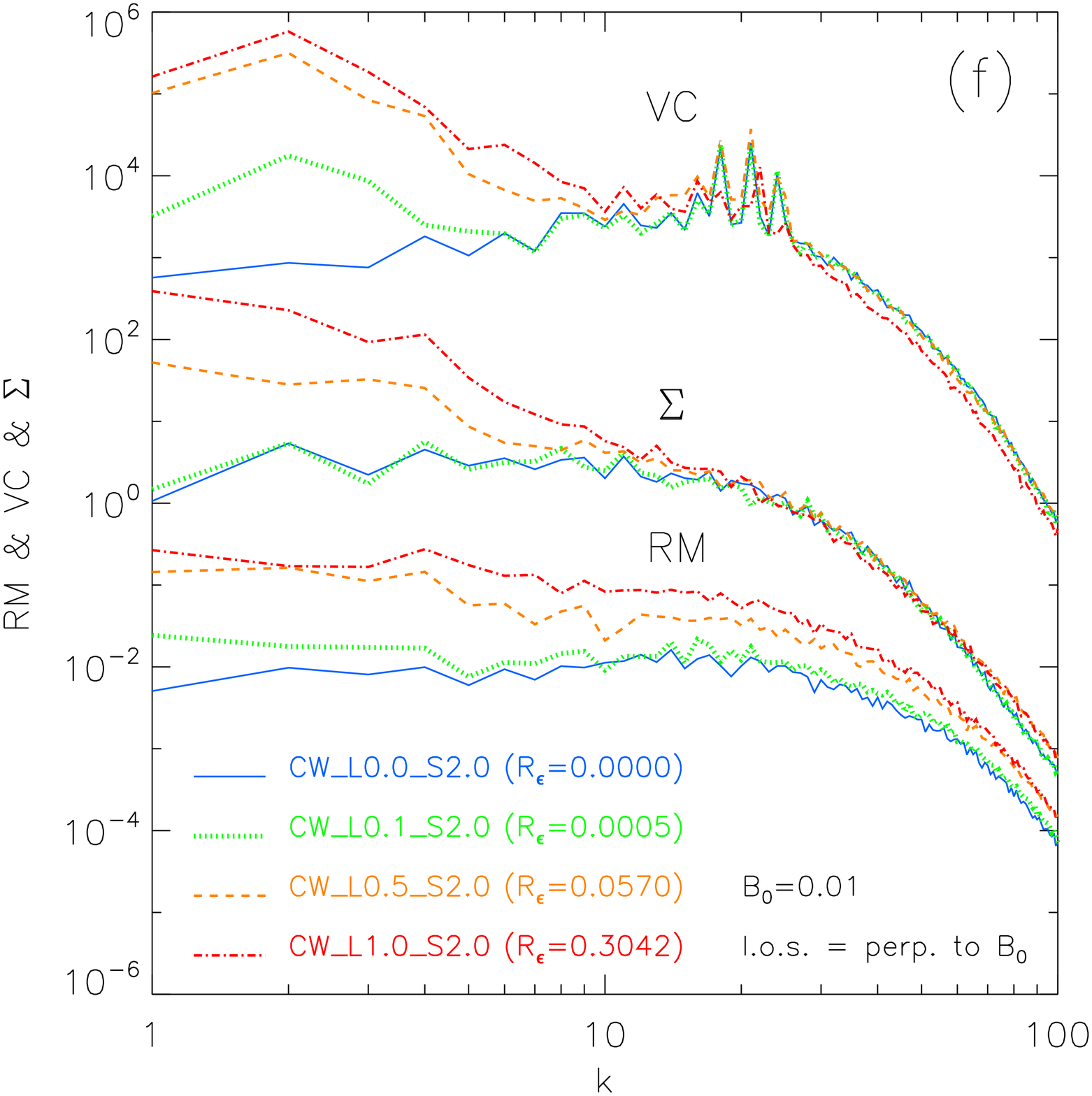} 
  \end{tabbing}
  \caption{
  \footnotesize{Compressible MHD turbulence simulations with a weak mean magnetic field. 
  (a) Time evolution of kinetic and magnetic energy densities. 
  (b) Time evolution of density fluctuations. 
  (c) Kinetic energy spectra. 
  (d) Magnetic energy spectra. 
  (e) Density spectra. 
  (f) Velocity centroids (Upper curves), Column densities (Middle curves), 
       and Rotation measures (Lower curves). 
      We use the same line conventions in all panels. 
      The sonic Mach number is $\sim$1 in all runs.   }
}
\label{f:dsp-weak}
\end{figure*}

\begin{figure*}
\begin{tabbing}
 \includegraphics[width=0.40\textwidth]{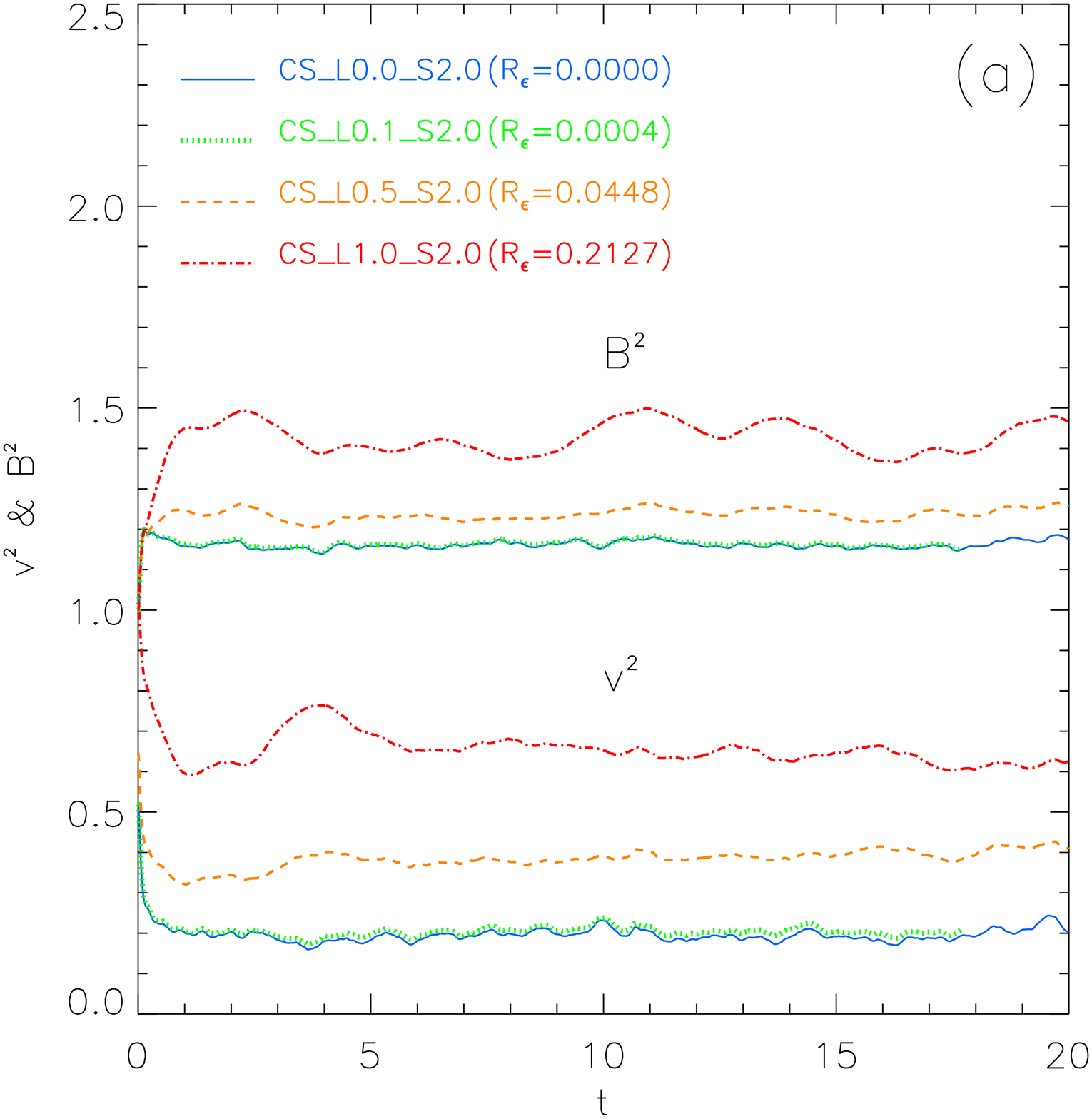}
 \=
  \includegraphics[width=0.40\textwidth]{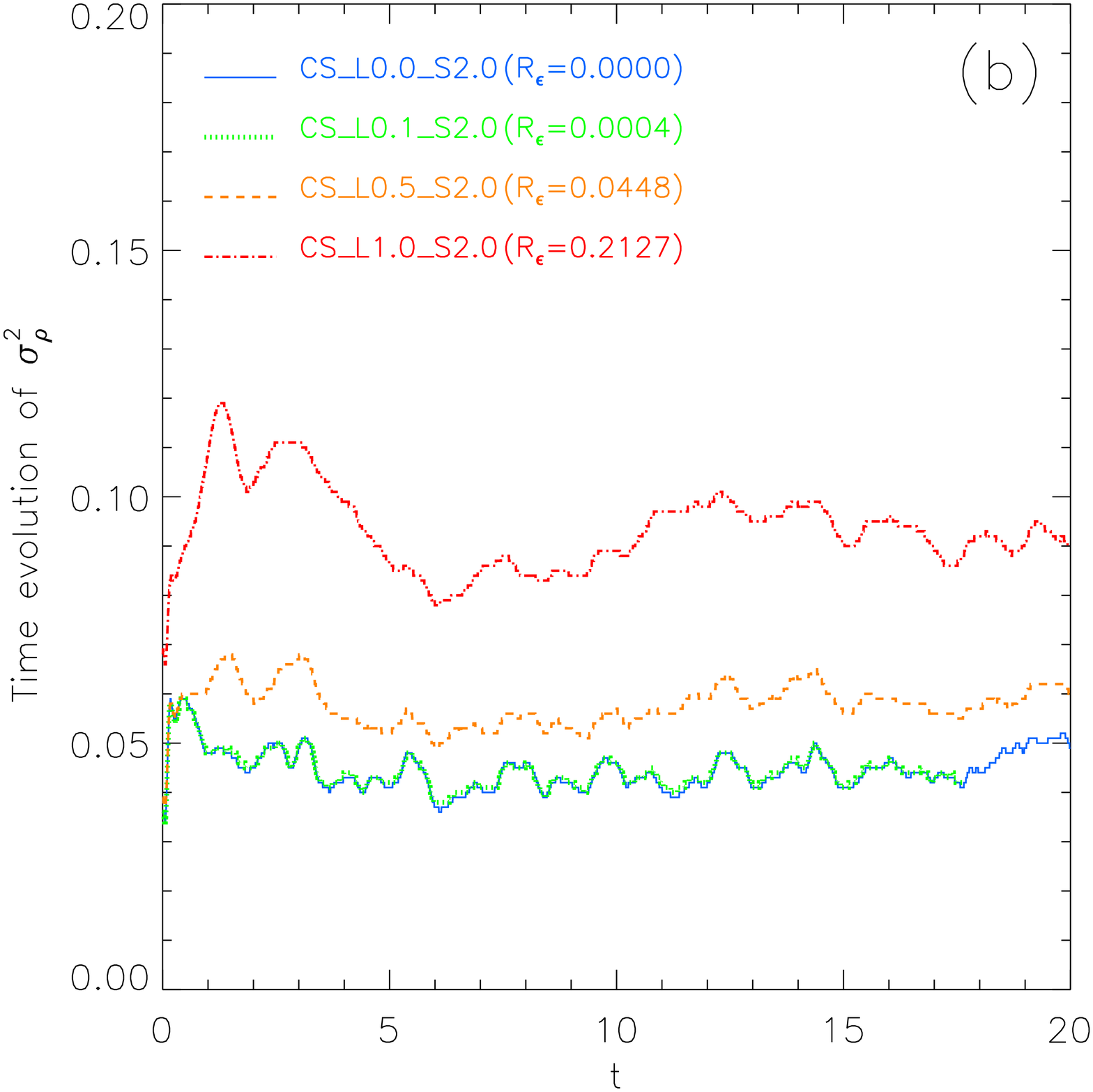} \\
  \includegraphics[width=0.40\textwidth]{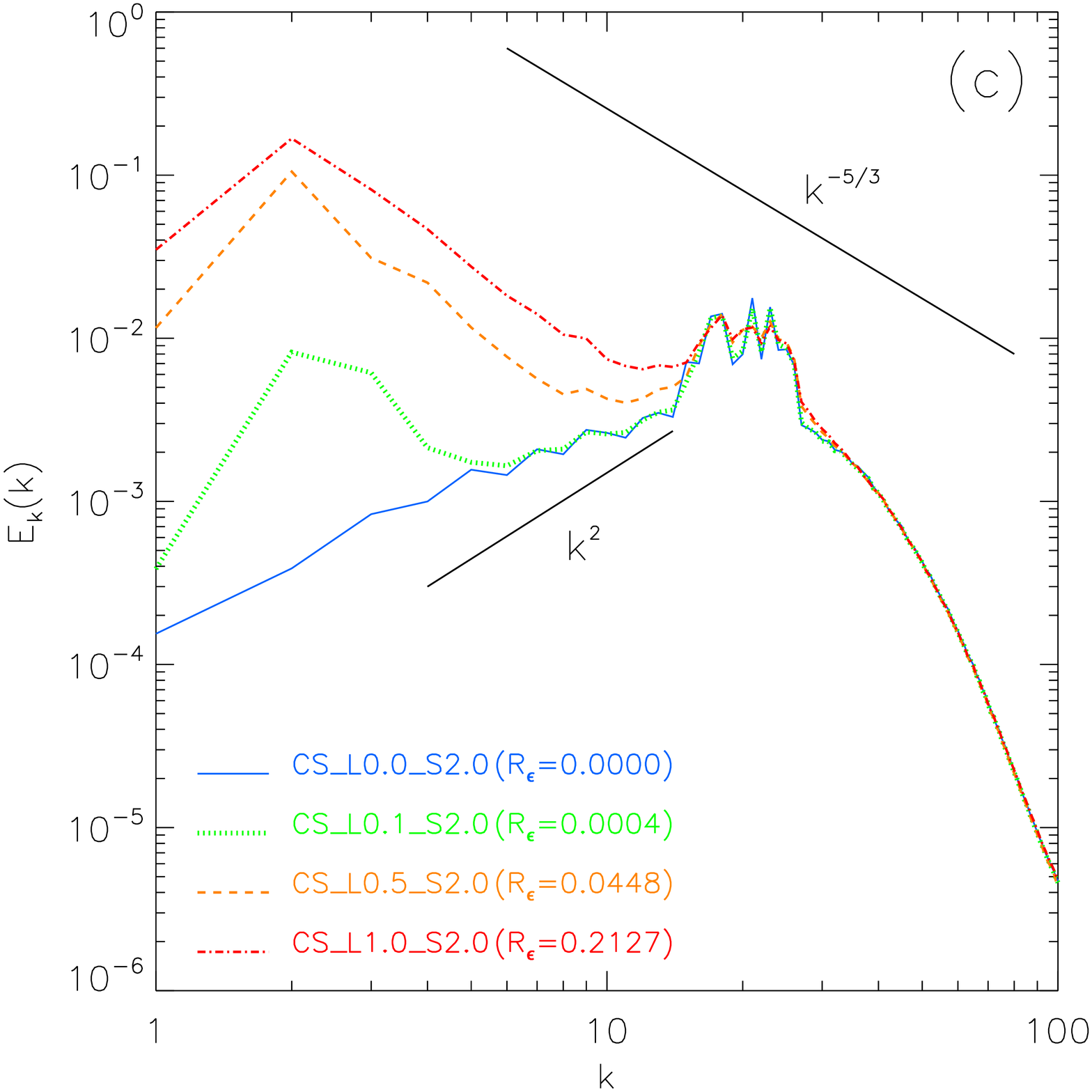}
  \>
  \includegraphics[width=0.40\textwidth]{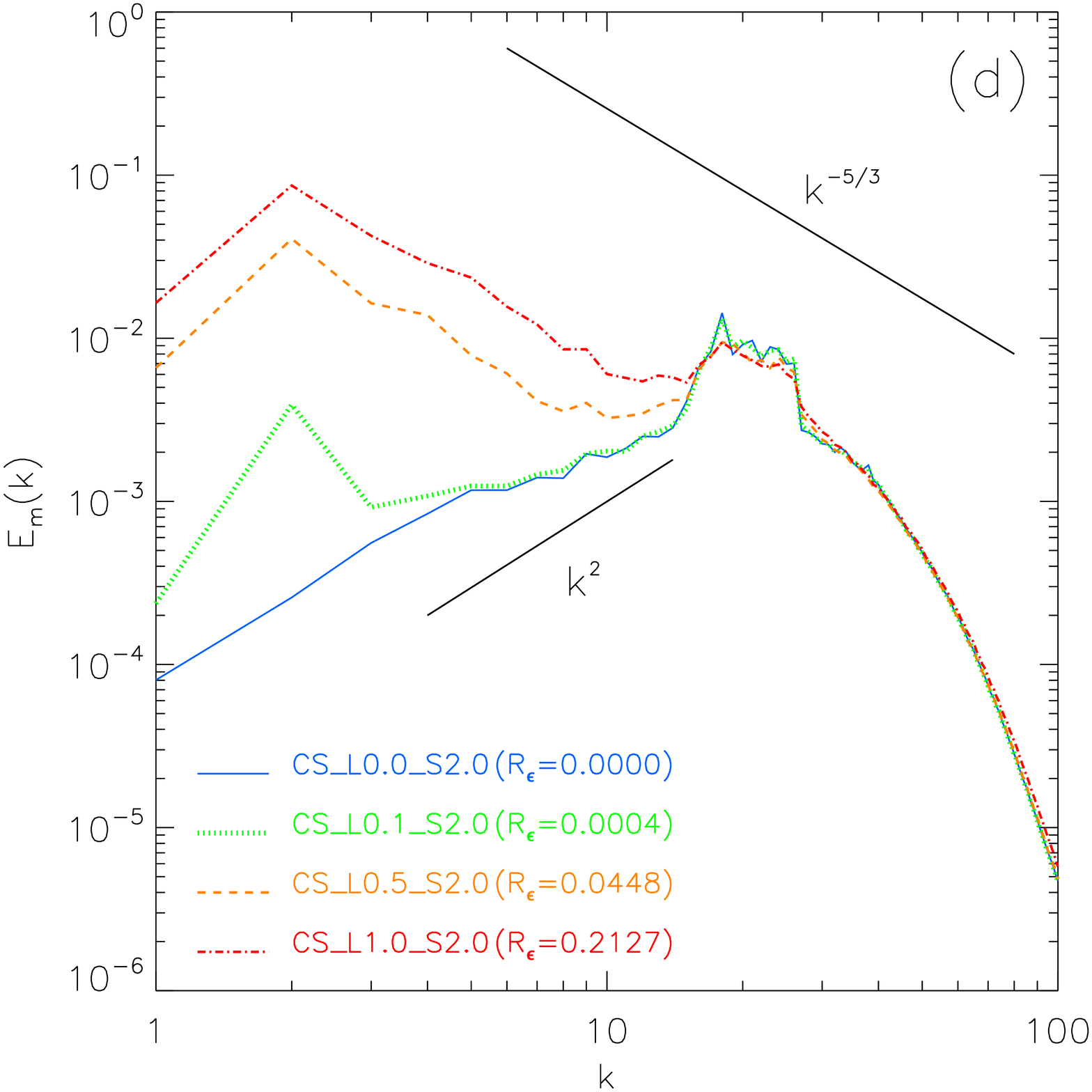} \\
  \includegraphics[width=0.40\textwidth]{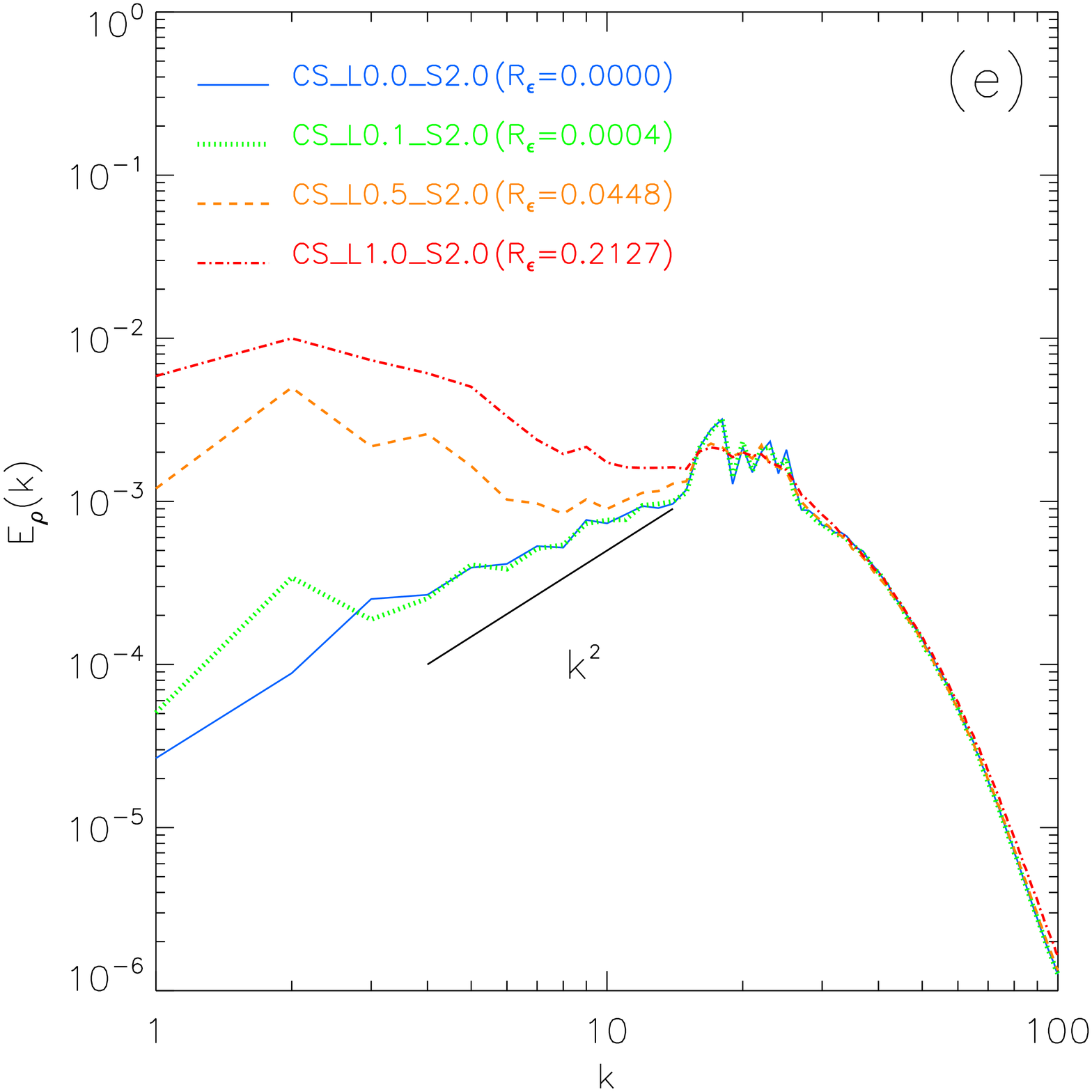}
  \>
  \includegraphics[width=0.40\textwidth]{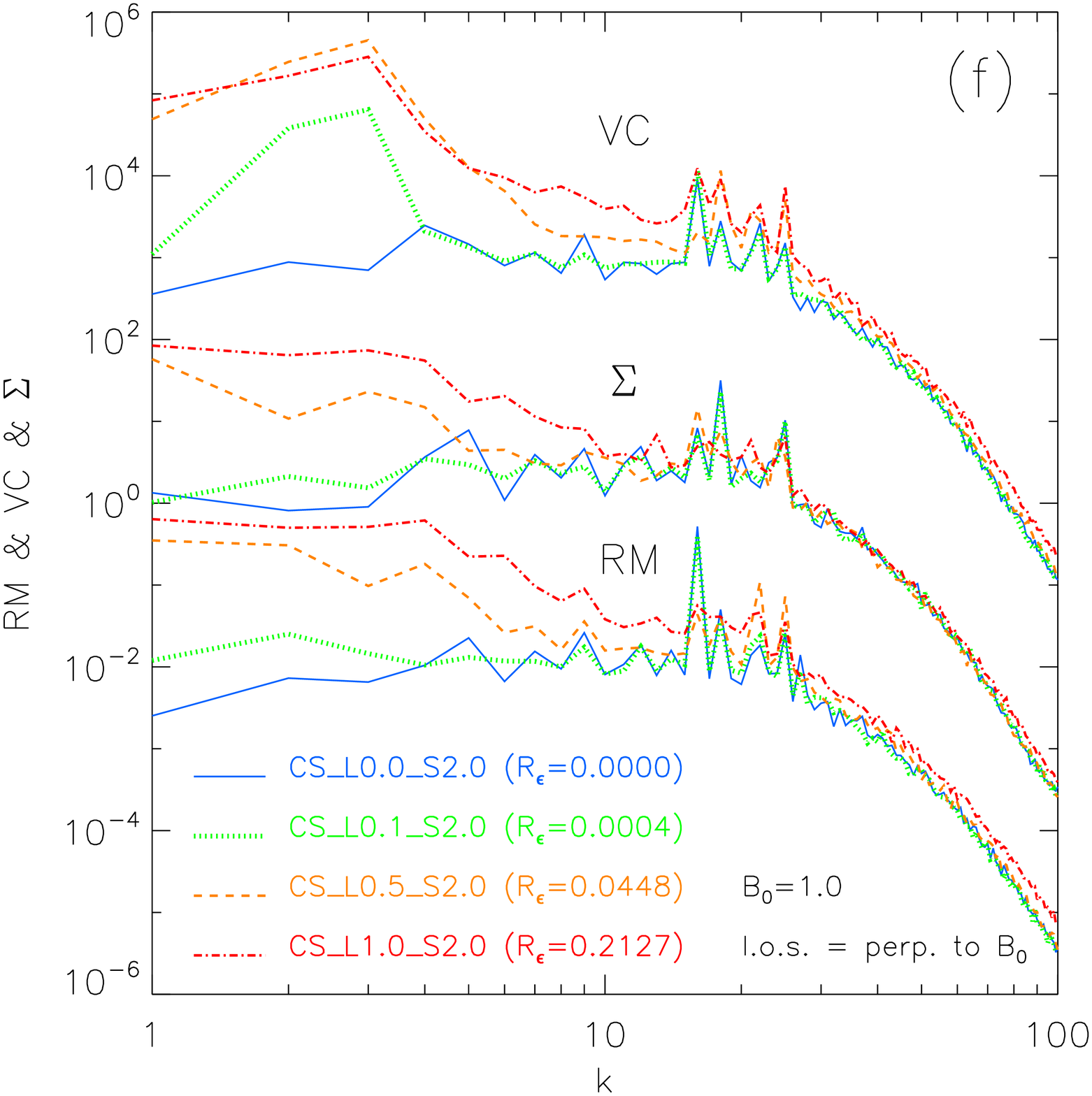} 
  \end{tabbing}
  \caption{
  \footnotesize{Compressible MHD turbulence simulations with a strong mean magnetic field. 
  (a) Time evolution of kinetic and magnetic energy densities.
  (b) Time evolution of density fluctuations. 
  (c) Kinetic energy spectra. 
  (d) Magnetic energy spectra. 
  (e) Density spectra. 
  (f) Velocity centroids (Upper curves), Column densities (Middle curves), 
      and Rotation measures (Lower curves).  
      We use the same line conventions in all panels. 
      The sonic Mach number is $\sim$1 in all runs.}
}
\label{f:dsp-str}
\end{figure*}

Multi-scale energy injection can also affect the density spectrum.
 In this work we focus on turbulence 
with a Mach number $M_S\sim 1.0$ and investigate how two-scale energy injection influences the density power spectrum\footnote{
    Kinetic and magnetic spectra in compressible runs are qualitatively similar to those in     
    incompressible runs (see panels (b) and (c) of Figures~\ref{f:dsp-weak} and \ref{f:dsp-str}).  
    Therefore, we do not pay much attention to the kinetic and magnetic spectra in this section. }.
In order to see the effect,
we use
the compressible MHD code described in Section \ref{sect:code2}.
We investigate density fluctuations in turbulence with either a weak ($B_{0}=0.01$) 
or a strong ($B_{0}=1.0$) mean magnetic field.
The simulation parameters are listed in Table 4.

Figure~\ref{f:dsp-weak} shows statistics of  turbulence with a weak  
mean magnetic field ($B_{0}=0.01$). 
Figure~\ref{f:dsp-weak}(a)  shows the time evolution of the kinetic and magnetic energy densities and 
Figure~\ref{f:dsp-weak}(b) shows that of the density fluctuations.
We start driven turbulence simulations without magnetic field 
at t=-3 and we introduce a weak mean field at t=0.
It seems that density fluctuations have already reached saturation at t$\sim$0.
This fact contrasts the time evolution of the magnetic fluctuations, which 
gradually increase and reach saturation after $t\sim 40$. 
The behaviors of the velocity and magnetic field are qualitatively 
similar to that of the incompressible runs (see Section \ref{sect:sd})\footnote{
  Note, however, that the strength of magnetic field at saturation  in compressible runs 
  is lower due to larger numerical diffusion. }.
The blue solid lines are for the run with a single driving at $k_S \sim$20 (i.e.~Run~$CW\_L0.0\_S2.0$).
When we inject additional energy into the large scale ($k_L\sim$2), the kinetic  energy density
increases.  When the large-scale energy-injection rate is small (for example, see the green dotted line; Run $CW\_L0.1\_S2.0$),
the resulting kinetic energy density is almost indistinguishable from that of the single-driving case (the blue solid line).
However, when the two injection rates are comparable 
(see, for example, the red dot-dashed line; Run $CW\_L1.0\_S2.0$),
the resulting kinetic energy density is highly boosted.

The density  fluctuations show a  trend similar to velocity. 
The green dotted line in Figure~\ref{f:dsp-weak}(b), 
which represents the variance of the density $\sigma_\rho^2$ for 
Run $CW\_L0.1\_S2.0$, almost coincides with the blue solid line.
However, the red dot-dashed line, which is for Run $CW\_L1.0\_S2.0$, is clearly
higher than the blue solid line.

Figures~\ref{f:dsp-weak}(c)$\sim$(e) show the kinetic, magnetic, and density spectra, respectively, 
at t$\sim$57.
In case of single-scale driving at $k_S\sim 20$ (the blue solid line), 
the slope of the kinetic and magnetic spectra for $k<k_S$
are compatible with $k^2$.
However, the slope of the density spectrum is flatter than $k^2$.
The overall behaviors of the kinetic and magnetic spectra are qualitatively similar to those of the incompressible simulations.

Let us focus on the density spectra (Figure~\ref{f:dsp-weak}(e)).
As we inject more energy at $k_L\sim 2$, the spectrum near $k_L\sim 2$ tends to go up.
When the additional large-scale energy injection is very small 
(see the green dotted  line; $R\epsilon \sim 0.0005$),
the spectrum at $k_L\sim 2$ shows only a very small increase, which might be due to
turbulence diffusion caused by motions near $l_S$.
When we further increase the large-scale energy-injection rate, 
the spectrum near $k_L\sim 2$ shows a notable increase 
(see the orange dashed or the red dot-dashed line; $R_\epsilon > 0.05$).
Even though the large-scale energy-injection rate is much smaller than 
the small-scale one,
the change of the spectrum is substantial. 
For example, the spectrum of Run $CW\_L0.5\_S2.0$ (the orange dashed line), in which
the large-scale energy-injection rate is only $\sim$6\% of the small-scale one, is notably higher than
that of Run $CW\_L0.0\_S2.0$ (the blue solid line) on large scales.
In Run $CW\_L1.0\_S2.0$ (the red dot-dashed line), 
in which the large-scale energy-injection rate is $\sim$30\% of the small-scale one,
the density spectrum at $k_L\sim 2$ is an order of magnitude larger than that at $k_S\sim 20$. 
The overall behavior of the density spectrum is similar to that of the kinetic spectrum.

Figure~\ref{f:dsp-str} shows statistics of  turbulence with a strong
mean magnetic field ($B_{0}=1.0$) . 
Except for the strength of the mean field, the other simulation setups are identical to those of the 
weak mean-field cases.
Figure~\ref{f:dsp-str}(a) shows the time evolution of the kinetic and magnetic energy densities 
and Figure~\ref{f:dsp-str}(b) shows that of the density fluctuations.
Since we introduce a strong mean field at t=0, 
the velocity fluctuations (see the lower curves in Figure~\ref{f:dsp-str}(a)) 
drop quickly and the magnetic fluctuations quickly increase at t$\sim$0.
It is interesting that the density fluctuations (see Figure~\ref{f:dsp-str}(b)) 
rise quickly as soon as we introduce a strong mean magnetic field at $t=0$.
The blue solid lines are for the run with a single driving at $k_S\sim$20.
When we inject additional energy into the large scale ($k_L\sim$2), 
the kinetic  and magnetic energy densities
increase. 
The overall behaviors of the velocity and magnetic fluctuations are similar to those in 
incompressible simulations.
The density  fluctuations also show a trend similar to the velocity or magnetic fluctuations. 
Note that the green dotted line in Figure~\ref{f:dsp-str}(b), 
which represents the variance of the density $\sigma_\rho^2$ for
Run $CS\_L0.1\_S2.0$ ($R_\epsilon \sim 0.0004$), 
almost coincides with the blue solid line (Run $CS\_L0.0\_S2.0$).

Figure~\ref{f:dsp-str}(c)$\sim$(e) show the kinetic, magnetic, and density spectra at t$\sim$17.
In the case of single-scale driving at $k_S\sim 20$ (the blue solid line), 
the slopes of the kinetic, magnetic, and density spectra for $k<k_S$ are roughly consistent with $k^2$.
As we inject additional energy at $k_L\sim 2$, the spectrum near $k_L\sim 2$ goes up.
The overall behavior of the density spectrum is similar to that of the kinetic spectrum.

As we have shown above, energy injections at two different scales affect not only the kinetic 
and magnetic spectra but also the density spectrum.  
It seems that density fluctuations are more tightly coupled to the velocity fluctuations
 than the magnetic fluctuations.
If the behavior of density fluctuations is indeed similar to that of velocity, we can 
apply the relations in Section \ref{sect:3.1} to density.
In this case, we have 
\begin{eqnarray}  
   \frac{ E_\rho (k_L) }{ E_\rho (k_S) } \sim \left( \frac{ \epsilon_L }{ \epsilon_S } \right)^{2/3}
                                  \left( \frac{ l_L }{ l_S } \right)^{5/3},    \label{eq:EE_rho} \\
   \frac{E_\rho (k_L)}{E_\rho (k_S)}\gtrsim 1, 
   \mbox{~~if~~} \frac{ \epsilon_L}{\epsilon_S} \gtrsim  \left( \frac{l_S}{l_L} \right)^{5/2} , \label{eq:EE_rho2} \\
   \frac{\sigma_{\rho,L}}{\sigma_{\rho,S}} \gtrsim 1, 
   \mbox{~~if~~}  \frac{ \epsilon_L}{\epsilon_S} \gtrsim \frac{l_S}{l_L}
\end{eqnarray}
(see Eqs.~(\ref{eq:EE}), (\ref{eq:EE_2}), and (\ref{eq:vv1}), respectively),
where the subscripts ``$\rho$'', ``L'', and ``S'' have their usual meanings.
{}Therefore, large-scale density fluctuations 
can be dominant even though the large-scale energy-injection rate is less than the small-scale one.

\subsection{Observational Implications}
In the previous subsection, we discussed how the density and other spectra are affected by the
energy injection at two scales.
It seems that the spectra deviate from those of a single-scale driving case even though 
a small fraction of the energy is injected into the large scale.
The density and other spectra in the previous subsection are for three-dimensional quantities.
Since what we observe are the quantities  projected onto the plane of sky, we briefly discuss in this subsection
the effects of multi-scale driving on column density ($\Sigma$), rotation measure (RM), 
and velocity centroids (VC):
\begin{eqnarray}
      \Sigma = \int \rho dy, \\
      RM=\int \rho B_y dy,   \\
      VC=\int \rho v_y dy,
\end{eqnarray}
where the summation is done along a direction perpendicular to the mean magnetic field ${\bf B}_0$.

Figure~\ref{f:dsp-weak}(f) shows the spectra of the projected quantities for the weak mean magnetic field cases.
For clarity, the VC spectra are multiplied by $300$ and the RM spectra are multiplied by 0.03.

In Figure~\ref{f:dsp-weak}(f), the column density spectra (i.e.,~the middle curves) 
look different from those of the 3D density field.
In fact, the spectrum of a 3D field ($E_{3D}$) and that of the projected field ($E_{proj}$) are related by
\begin{equation}
      E_{proj}(k) \propto E_{3D}(k)/k.
\end{equation}
In case of small-scale driving only (the blue solid line), we can see a very broad flat spectrum for $k\lesssim k_S$
in the column density spectrum. 
This is consistent with the fact that the spectrum of the 3D density field is a bit flatter than $k^{2}$ 
(see the blue solid line in Figure \ref{f:dsp-weak}(e)).
When a small fraction of energy is added at $k\sim k_L$, the average
 slope of the column density spectrum between $k_L$ and $k_S$ decreases.
 Even a small addition of energy on a large scale can make a big difference
in the column density spectrum between $k_L$ and $k_S$.
If the density fluctuations follow the relation in Eq.~(\ref{eq:EE_rho}), we have
\begin{equation} 
     \frac{ E_\Sigma (k_L) }{ E_\Sigma (k_S) } 
     \sim \frac{ E_\rho (k_L) k_S }{ E_\rho (k_S)k_L }
     \sim
      \left( \frac{ \epsilon_L }{ \epsilon_S } \right)^{2/3}
                                  \left( \frac{ l_L }{ l_S } \right)^{8/3}.   
\end{equation}
Therefore, we expect that, if $R_\epsilon > (l_S/l_L)^4 $, the average slope of 
column density spectrum between $k_L$ and $k_S$ becomes negative.

In Figure~\ref{f:dsp-weak}(f),
the average slope of the VC spectrum  for the case of small-scale driving only 
(i.e.,~the blue solid line among the upper curves)
is slightly positive for $k<k_S$ and, as in the column density spectrum,
the spectrum for $k<k_S$ changes as we inject additional energy at $k\sim k_L$.
It seems that the effects of two-scale driving is more pronounced in the VC spectrum, which
implies that the VC observation is a good way to detect imprints of multi-scale energy injections.
Note, however, that the VC spectrum reflects velocity statistics for subsonic or transonic turbulence and
density statistics for supersonic turbulence (Esquivel \& Lazarian 2005).
Therefore, the VC observation would be better suited for subsonic or transonic turbulence as in our simulations.

In Figure~\ref{f:dsp-weak}(f), the behavior of the RM spectra (i.e.,~the lower curves)
is clearly different from
that of the column density or the VC.
The RM spectra are not strongly affected by multi-scale driving.
The overall behavior of the RM spectra seems to be similar to that of magnetic field spectra.
Therefore it may be more difficult to detect multi-scale driving using the RM observations
when the mean field is weak.

Figure~\ref{f:dsp-str}(f) shows spectra of the projected quantities for the strong mean magnetic field cases.
For clarity, the VC spectra are multiplied by 300 and the RM spectra are multiplied by 0.003.
Unlike the cases with a weak mean magnetic field, all three projected quantities behave similarly.
Note that even the RM spectra show multi-scale driving effects.
Therefore, when the mean field is strong, it may not be difficult to detect multi-scale driving effects.

\section{Discussion}

We have performed numerical simulations of incompressible/compressible MHD turbulence with
either single or double energy-injection scales. 
In real astrophysical fluids, 
there are many possible energy sources acting on different scales.
Therefore,
our results can be applied to many astrophysical fluids. 

\subsection{Turbulence in the ICM}
Models of turbulence permeated by a weak mean magnetic field (see Section \ref{sect:sd})
may correspond to ICM turbulence. 
Suppose that there is a dominant small-scale energy-injection mechanism,
such as jets from an active galactic nucleus (AGN), and
an additional large-scale energy-injection mechanism, such as cosmological shocks.
The characteristic length scale of the latter can be comparable to $\sim 1Mpc$, while   
that of the former can be an order of magnitude smaller.
In this case, since the scale separation of the two energy-injection mechanisms is of order $\sim10$,
we may be able to observe two comparable peaks in the kinetic energy spectrum
even if the contribution of the large-scale energy injection is much smaller 
than that of the small-scale one.
Indeed, our results show that
we can see two peaks of similar heights in the kinetic energy spectrum even though 
the large-scale energy-injection rate is hundreds of times smaller than the small-scale counterpart 
(see   Eq.~(\ref{eq:18});  
see  also  Figure \ref{f:weakB}(c)). 
However, it seems that, even if there are two comparable peaks in kinetic energy spectrum, 
it may not be easy to observe 
two peaks in the spectra of projected quantities, such as column density or RM (see Figure~\ref{f:dsp-weak}(f)).
Nevertheless, we may be able to observe a shallow or flat spectrum for $k_L<k<k_S$ in those spectra.

It is also possible that dominant energy is injected on large scales in the ICM.
In this case, as we discussed in Section \ref{sect:ld} (see Eq.~(\ref{eq:9})), 
the small-scale driving should be as energetic
as the large-scale one in order to produce  a visible bump at the small-scale in
kinetic energy spectrum. 

\subsection{Turbulence in the ISM or in the Solar Wind}
Models of turbulence permeated by a strong mean magnetic field (see Section \ref{sect:strongB0}) 
may correspond to
Galactic turbulence 
or solar wind turbulence.
In Galactic turbulence, supernova explosions can be dominant sources of energy,
which inject energy on scales of tens of parsecs.

In addition, a larger-scale driving, e.g., magnetorotational instabilities, may exist.
If the larger-scale driving operates on scales an order of magnitude larger than the supernova,
we will be able to see two distinct peaks in the energy spectra even though
the large-scale energy-injection rate is hundreds of times smaller than the energy-injection rate of
supernova explosions.
Since the kinetic and magnetic spectra behave similarly in turbulence with a strong mean field,
this statement is applicable to both kinetic and magnetic spectra.
Indeed,
our turbulence model `Run $IS\_L0.1\_S5.0$' 
(the blue dotted line in Figure \ref{f:strongB}; 
see also Run $CS\_L0.1\_S2.0$, the green dotted line,  in   Figure~\ref{f:dsp-str}) 
implies that
the large-scale peak appears even though the 
ratio of the large-scale and the supernova-scale energy-injection rates $R_\epsilon$ is as small as $\sim 10^{-3}$. 
Therefore, if we are able to observe the kinetic or magnetic energy spectrum of  ISM turbulence for
a wide range of scales up to hundreds of parsecs, 
we may be able to observe two peaks in the spectrum. 

On the other hand, in addition to dominant supernova explosions,
sub-dominant smaller-scale driving sources (e.g. stellar winds or outflows) may exist.
In this case, it may not be easy to observe the effects of the smaller-scale sources in
the kinetic or magnetic energy spectrum, unless
they inject energy
comparable to that of the supernova explosions.
When they inject as much energy as supernova explosions, we can see the effects in the column density and RM
spectra.
Run $CS\_L1.0\_S2.0$ (see Figure~\ref{f:dsp-str}), 
in which the small-scale energy-injection rate is about 5 times larger than
the large-scale one, shows that the average density spectrum between $k_L$ and $k_S$
is flatter than a Kolmogorov spectrum.
Therefore, when $R_\epsilon 1$, we expect to see power spectra of column density and RM
flatter than a Kolmogorov spectrum.
The flat second-order structure function observed by Haverkorn et al. (2008) might 
be caused by simultaneous energy injections on small and large scales.

\subsection{Field-line Divergence and Turbulence Diffusion}
When there are multi-scale energy injections, many physical properties of turbulence can change.
For example, magnetic field-line divergence and related particle transport properties 
(e.g.~thermal diffusion), the turbulence diffusion coefficient, and the growth rate of a localized
seed magnetic field (see Cho \& Yoo 2012; Cho 2013) will change in the presence
of multi-scale energy injections.

Let us consider magnetic field-line divergence first.
Suppose that the mean magnetic  field is strong and that we have a single-scale energy injection at $k=k_S$.
Then, roughly speaking, the distance between two neighboring magnetic field lines 
on scales larger than $l_S ~(\sim1/k_S)$ will 
show a diffusive behavior. Magnetic field-line divergence in this case will follow
\begin{equation}
  <y^2>^{1/2} \sim l_S (b_S/B_0) (z/l_S)^{1/2},
\end{equation}
where $y$ is the separation between two field lines and $z$ is the distance along
the mean magnetic field direction.
If we inject an additional energy on a larger scale $l_L$ ($> l_S$), 
we expect a superdiffusion of magnetic field lines 
between the two driving scales.
Note that even a tiny amount of large-scale energy injection will suffice to make a dramatic  change  
in magnetic field-line divergence if the two driving scales
are well separated.
When $z\sim l_L$, field-line divergence by the large-scale driving will be
\begin{equation}
 \sim l_L \frac{b_L}{B_0}\sim \frac{\epsilon_L^{1/3}}{B_0} l_L^{4/3}
    \sim l_L^{1/2} \frac{ E(k_L)^{1/2} }{B_0},
\end{equation}
while that by the small-scale driving will be
\begin{eqnarray}
 \sim l_S \frac{b_S}{B_0}\left( \frac{l_L}{l_S} \right)^{1/2}
       \sim \frac{\epsilon_S^{1/3}}{B_0} l_S^{4/3} \left( \frac{l_L}{l_S} \right)^{1/2} \nonumber \\
       \sim l_S^{1/2} \frac{ E(k_S)^{1/2}}{B_0} \left( \frac{l_L}{l_S} \right)^{1/2}\,
\end{eqnarray}
where we use $\epsilon \sim b^3/l$ and $b \sim \sqrt{ kE(k) }$.
Field-line divergence by the large-scale driving dominates if
\begin{equation}
 \frac{\epsilon_L}{\epsilon_S} \gtrsim \left( \frac{ l_S }{ l_L } \right)^{5/2},
\end{equation}
or
\begin{equation}
   \frac{ E(k_L) }{ E(k_S) } \gtrsim 1.
\end{equation}
Therefore, even a tiny amount of energy injection at the large scale 
can significantly alter the rate of field-line divergence and also
the rate of particle transport.

Turbulence diffusion is also affected by two-scale driving.
Suppose that we have fully developed turbulence driven at $l_S$.
If we inject additional energy on a scale  larger than $l_S$,   
the turbulence diffusion coefficient can be also greatly enhanced by the large-scale driving.
The turbulence diffusion coefficient is $\sim l v$.
Therefore, turbulence diffusion by the large-scale motions will dominate
when
\begin{equation}
  \frac{ l_L (\epsilon_L l_L)^{1/3} }{ l_S (\epsilon_S l_S)^{1/3} } \gtrsim 1 \mbox{~~or~~}
  \frac{ \epsilon_L }{\epsilon_S } \gtrsim \left(  \frac{ l_S }{ l_L } \right)^{4}.
\end{equation}

\subsection{Turbulence Dynamo}  \label{sect:turb_dyn}
The velocity field can stretch the magnetic field lines and transfer kinetic energy to the magnetic field. 
This small-scale turbulence dynamo is so efficient 
that $b^2$ can grow up to a value only slightly smaller than $v^2$
(e.g.~Cho \& Vishniac 2000). 
If we have a driving on a single scale, 
the growth of a weak \textit{uniform} magnetic field in a turbulent medium occurs in three stages
(see  Schl\"uter \& Biermann 1950;  Cho \& Vishniac 2000; 
       Schekochihin \& Cowley 2007; Cho et al. 2009; Beresnyak 2012): 
(1) exponential growth: stretching of the
magnetic field lines occurs most actively near the velocity dissipation
scale first, and the magnetic energy grows exponentially. 
(2) linear growth: the exponential growth stage ends when the magnetic energy
becomes comparable to the kinetic energy at the dissipation
scale. The subsequent stage is characterized by a linear growth
of magnetic energy and a gradual increase of the stretching scale.
(3) saturation: the amplification of the magnetic
field stops when the magnetic energy density becomes comparable
to the kinetic energy density.
In Figure~\ref{f:weakB}(b), the lower purple solid line 
 shows the time evolution of the magnetic field 
in the run with single driving 
(Run  $IW\_L1.0\_S0.0$).  
Although it is not clearly visible in the figure, the exponential stage occurs before $t\sim 5$.
The linear growth stage happens between $t\sim 5$ and $\sim 35$.

When turbulence is driven at two different scales, 
the saturation level of $b^2$ shows a sudden change as $R_\epsilon~(= \epsilon_L/\epsilon_S)$ increases.
Figure \ref{f:weakB}(a) shows that the saturation level of
$b^2$ does not change much when  $R_\epsilon  \lesssim 0.1$.
When $R_\epsilon \gtrsim 0.1$,
the saturation level of $b^2$ goes up as $R_\epsilon$ goes up.
On the other hand, Figure \ref{f:weakB}(b) shows that, when $R_\epsilon \gtrsim 0.4$,
 the final saturation level of $b^2$ 
is almost same for all runs.

Turbulence diffusion by the motions near $l_S$ could explain
such a behavior of $b^2$.
When turbulence is driven at two scales, $l_L$ and $l_S$, 
two competing effects
determine the fate of the large-scale magnetic field $b_L$.
First, turbulence motions near $l_L$ stretch the magnetic field lines and 
amplify the large-scale magnetic energy density at a rate $\sim C v_L b_L^2/l_L$, where
$C$ is a small number.
Second, turbulence motions near $l_S$ provide
turbulence diffusion and destroy large-scale magnetic energy density at a rate
 $\sim (l_S v_S) b_L^2/l_L^2$.
Therefore, we will have a growth of the large-scale magnetic energy density $b_L^2$ if
\begin{equation}
 \frac{v_L}{v_S} \gtrsim \frac{l_S}{l_L} C^{-1},  \label{eq:dynamo_v}
\end{equation}
or, from Eq.~(\ref{eq:vv}),
\begin{equation}
    \frac{ \epsilon_L }{ \epsilon_S } \gtrsim \left( \frac{ l_S }{ l_L } \right)^4 C^{-3}.
    \label{eq:dynamo_e}
\end{equation}

Our results imply that $(l_S/l_L) C^{-1} \sim O(1)$ in our simulations.
In this case, when $\epsilon_L/\epsilon_S \lesssim 0.1$,
turbulence dynamo is not possible at the scale near $ l_L$
and the saturation level
of the magnetic energy density $b^2$
is mainly determined by $v_S^2$.
When $\epsilon_L/\epsilon_S$ is larger than a few times $0.1$,
the turbulence diffusion effect becomes clearly sub-dominant and 
the large-scale $l_L$ becomes the main scale for the turbulence dynamo.
The saturation level of $b^2$ in this case will be $\sim v_L^2$.
If $(l_L/l_S) \gg 10$, $(l_S/l_L) C^{-1}$ will be much smaller than 1.
In this case, when  $(v_L/v_S) \gtrsim (l_S/l_L) C^{-1}$, 
the saturation level of $b^2$ will be $\sim v_L^2+v_S^2$, which will be $\sim v_L^2$
when $(v_L/v_S) > 1$.

\subsection{Effects of Magnetic Prandtl Number and Mach Number}
In this paper, 
we have considered fluids with unit magnetic Prandtl number ($\nu/\eta$) only.
  However, in many astrophysical fluids, viscosity is not negligibly small, while magnetic
  diffusivity is very small, so that the 
  magnetic Prandtl number is much larger than unity (i.e.~$\nu \gg \eta$).
  In the ICM, for example, the Reynolds number $Lv/\nu$, where $L$
is the outer scale of turbulence and $v$ is the rms velocity, 
is less than $\sim 10^{3}$ if we use the Spitzer (1962)
formula for the viscosity, but the magnetic Reynolds number $Lv/\eta$
is much larger than $10^3$.
In this case, the viscous-cutoff wavenumber $k_d$, at which 
velocity spectrum drops quickly due to viscous damping, is not much larger than the
driving wavenumber.

In high magnetic Prandtl number fluids, it is relatively easy to
understand the behavior of the velocity spectrum in the presence of driving at two scales.
We expect that
the relative magnitude of two wavenumbers, the  wavenumber for small-scale forcing $k_S$ and 
the viscous-cutoff wavenumber for large-scale driving $k_{d,L}$, may matter.
If $k_S \gg k_{d,L}$, we will see two disjointed velocity spectra:
a spectrum with a peak at $k=k_L$ and a viscous cutoff at $k=k_{d,L}$ and another
spectrum with  a peak at $k=k_S$ and a viscous cutoff at $k=k_{d,S}$.
If $k_S < k_{d,L}$, we will see a connected velocity spectrum with two peaks at $k_L$ and $k_S$ and a viscous cutoff
at $\sim k_{d,S}$.
However, the behavior of the magnetic spectrum seems to be very complicated and
further numerical investigations will be necessary.


In compressible simulations (Section \ref{sect:4}), 
we have considered fluids with unit Mach number ($v/C_s$) only, which
may be acceptable for the ICM or the solar wind.
However, in some astrophysical fluids, e.g.,~molecular clouds, the Mach number
can be much larger than unity. 
Since the Mach number affects velocity and density spectra (Beresnyak et al. 2005;
Kim \& Ryu 2005),
as well as turbulence dynamo efficiency (Haugen, Brandenburg, \& Mee 2004), 
 further numerical studies will be necessary. We will address all these issues elsewhere.

\section{Conclusion and summary}

In this paper, we have studied the statistical properties of weakly and strongly magnetized  
incompressible/compressible
 turbulence driven by solenoidal forcing in two wavenumber ranges in Fourier space. 
 We have found that driving turbulence at two different scales
  affects kinetic, magnetic, and density spectra differently.
  
  When turbulence is driven only at a small scale $l_S$, all spectra have peaks
      near the driving scale, regardless of the strength of the mean magnetic field
      (see the solid lines in Figures~2(c) and (e), Figures~4(b) and (c), 
                              5(c)-(e), and 6(c)-(e)).
      When we inject additional energy at a larger scale $l_L$, 
      the spectra of velocity, magnetic field, and density change.
      The results we have found from this work imply the following changes.

\begin{enumerate}
\item{Velocity spectra: 
         when $\epsilon_L/\epsilon_S$ is less than $\sim 1$, we can see two peaks in the spectrum:
         one at $k_L ~(\propto 1/l_L)$ and the other at $k_S ~(\propto 1/l_S)$.
         Here $\epsilon_L$ and $\epsilon_S$ are the energy-injection rates.
         The relation between $E(k_L)/E(k_S)$ and $\epsilon_L/\epsilon_S$ 
         in runs with either a weak (Figure~3) or a strong (Figure~4(d)) mean magnetic field
         follows the theoretical expectation for hydrodynamic turbulence closely
         (Equation~(\ref{eq:EE})).
         According to Equation~(\ref{eq:EE}), even a tiny amount of energy injection
         at $l_L$ will suffice to make $E(k_L) > E(k_S)$ if $l_L \gg l_S$ (see Equation~(\ref{eq:EE_2})).
         }
         
\item{Magnetic spectra:          
         when the mean magnetic field is weak, the magnetic spectrum does not show two distinct peaks
         in our simulations, in which $k_S/k_L \sim 10$.
         Instead, we observe a monotonic change in the spectrum between $k_L$ and $k_S$.
         The average slope of the magnetic energy spectrum between the two driving scales
        tends to increase as we inject more energy into the large scale.
        The relation between $E(k_L)/E(k_S)$ and $\epsilon_L/\epsilon_S$ 
         does not match the theoretical expectation for hydrodynamic turbulence
         (Equation~(\ref{eq:EE})).
         
         When the mean magnetic field is strong, the behavior of the magnetic spectrum
         is very similar to that of the kinetic spectrum.
         Therefore, the relation between $E(k_L)/E(k_S)$ and $\epsilon_L/\epsilon_S$ 
         for the magnetic field
         also follows the theoretical expectation given in
         Equation~(\ref{eq:EE}) closely.
         }
 
\item{Density spectra: 
        the behavior of the density spectrum is more or less similar to that of the kinetic spectrum.
        We have also observed change in the column density, rotation measure, and
        velocity-centroid spectra in the presence of energy injections at two scales.
        }

\end{enumerate}

The turbulence dynamo is also affected by driving at two scales.
When $(v_L/v_S) \lesssim (l_S/l_L) C^{-1}$, where $C$ is a small number,
the small scale $l_S$ is the main
scale for turbulence dynamo and we have $b^2 \sim v_S^2$ during saturation.
On the other hand, when $(v_L/v_S) \gtrsim (l_S/l_L) C^{-1}$,
the turbulence dynamo at
 the large scale $l_L$ becomes possible and
we will have  $b^2 \sim v_S^2 + v_L^2$ during saturation.

\acknowledgements
This research was supported by the National R \& D Program through 
the National Research Foundation of Korea (NRF) 
funded by the Ministry of Education (No. 2011-0012081).
The initial calculations for the hydrodynamic turbulence were started with the help of 
Youngdae Lee, whom we would like to thank.


\clearpage

\begin{deluxetable}{ccccccccc}
\tablewidth{0pt}
\tablecaption{Weakly-magnetized incompressible turbulence simulations with a fixed small-scale driving}
\tablehead{\colhead{Model} & \colhead{$f_{L}$}\tablenotemark{a} & \colhead{$f_{S}$}\tablenotemark{b} & \colhead{$\epsilon_{T}$}\tablenotemark{c} & \colhead{$\epsilon_{L}$}\tablenotemark{d} & \colhead{$\epsilon_{S}$}\tablenotemark{e} & \colhead{$t_{1}$} & \colhead{$t_{2}$}\tablenotemark{f} & \colhead{$R_{\epsilon}$}\tablenotemark{g}}
\startdata
$IW\_L0.0\_S2.0$ & 0.0 & 2.0 & 0.38498 & 0.00000 & 0.38498 & 10 & 70 & 0.00000\\
$IW\_L0.1\_S2.0$ & 0.1 & 2.0 & 0.37770 & 0.00145 & 0.37625 & 10 & 75 & 0.00386\\
$IW\_L0.5\_S2.0$ & 0.5 & 2.0 & 0.33058 & 0.02761 & 0.30297 & 10 & 85 & 0.09112\\
$IW\_L0.7\_S2.0$ & 0.7 & 2.0 & 0.31874 & 0.04778 & 0.27096 & 10 & 100 & 0.17633\\
$IW\_L0.8\_S2.0$ & 0.8 & 2.0 & 0.31784 & 0.06070 & 0.25714 & 10 & 100 & 0.23604\\
$IW\_L1.0\_S2.0$ & 1.0 & 2.0 & 0.32362 & 0.08755 & 0.23607 & 10 & 65 &  0.37087
\enddata
\tablenotetext{a}{Amplitude of large-scale (2$\leq$k$\leq$$\sqrt12$) forcing in code units}
\tablenotetext{b}{Amplitude of small-scale (15$\lesssim $k$\lesssim $26) forcing in code units}
\tablenotetext{c}{Total energy-injection rate (=$\epsilon_L + \epsilon_S$)}
\tablenotetext{d}{Large-scale energy-injection rate}
\tablenotetext{e}{Small-scale energy-injection rate}
\tablenotetext{f}{energy-injection rates are averaged from $t_{1}$ to $t_{2}$}
\tablenotetext{g}{$R\epsilon \equiv \epsilon_{L}  /  \epsilon_{S}$}
\end{deluxetable}

\begin{deluxetable}{ccccccccc}
\tablewidth{0pt}
\tablecaption{Weakly-magnetized  incompressible turbulence simulations with a fixed large-scale driving}
\tablehead{\colhead{Model} & \colhead{$f_{L}$}\tablenotemark{a} & \colhead{$f_{S}$}\tablenotemark{b} & \colhead{$\epsilon_{T}$}\tablenotemark{c} & \colhead{$\epsilon_{L}$}\tablenotemark{d} & \colhead{$\epsilon_{S}$}\tablenotemark{e} & \colhead{$t_{1}$} & \colhead{$t_{2}$}\tablenotemark{f} & \colhead{$R_{\epsilon}$}\tablenotemark{g}}
\startdata
$IW\_L1.0\_S0.0$ & 1.0 & 0.0 & 0.08903 & 0.08903 & 0.00000 & 10 & 70 & -\\
$IW\_L1.0\_S0.5$ & 1.0 & 0.5 & 0.10419 & 0.08860 & 0.01559 & 10 & 65 & 5.68300\\
$IW\_L1.0\_S1.0$ & 1.0 & 1.0 & 0.14995 & 0.08873 & 0.06122 & 10 & 65 & 1.44941\\
$IW\_L1.0\_S2.0$ & 1.0 & 2.0 & 0.32362 & 0.08755 & 0.23607 & 10 & 65 & 0.37087 
\enddata
\tablenotetext{a-g}{Same as Table 1}
\end{deluxetable}

\begin{deluxetable}{ccccccccc}
\tablewidth{0pt}
\tablecaption{Strongly-magnetized incompressible turbulence simulations with a fixed small-scale driving}
\tablehead{\colhead{Model} & \colhead{$f_{L}$}\tablenotemark{a} & \colhead{$f_{S}$}\tablenotemark{b} & \colhead{$\epsilon_{T}$}\tablenotemark{c} & \colhead{$\epsilon_{L}$}\tablenotemark{d} & \colhead{$\epsilon_{S}$}\tablenotemark{e} & \colhead{$t_{1}$} & \colhead{$t_{2}$}\tablenotemark{f} & \colhead{$R_{\epsilon}$}\tablenotemark{g}}
\startdata
$IS\_L0.0\_S5.0$ & 0.0 & 5.0 & 0.97557 & 0.00000 & 0.97557 & 10 & 100 & 0.00000\\ 
$IS\_L0.1\_S5.0$ & 0.1 & 5.0 & 0.99199 & 0.00044 & 0.99156 & 10 & 36 & 0.00044\\
$IS\_L0.5\_S5.0$ & 0.5 & 5.0 & 0.97536 & 0.01106 & 0.96430 & 10 & 30 & 0.01147\\
$IS\_L0.8\_S5.0$ & 0.8 & 5.0 & 0.97248 & 0.02823 & 0.94425 & 10 & 70 & 0.02989\\
$IS\_L1.0\_S5.0$ & 1.0 & 5.0 & 0.97923 & 0.04357 & 0.93567 & 10 & 35 & 0.04656
\enddata
\tablenotetext{a-g}{Same as Table 1}
\end{deluxetable}

\begin{deluxetable}{ccccccccc}
\tablewidth{0pt}
\tablecaption{Weakly- and strongly-magnetized compressible turbulence simulations with a fixed small-scale driving}
\tablehead{\colhead{Model} & \colhead{$f_{L}$}\tablenotemark{a} & \colhead{$f_{S}$}\tablenotemark{b} & \colhead{$\epsilon_{T}$}\tablenotemark{c} & \colhead{$\epsilon_{L}$}\tablenotemark{d} & \colhead{$\epsilon_{S}$}\tablenotemark{e} & \colhead{$t_{1}$} & \colhead{$t_{2}$}\tablenotemark{f} & \colhead{$R_{\epsilon}$}\tablenotemark{g}}
\startdata
$CW\_L0.0\_S2.0 $ & 0.0 & 2.0 & 0.37586 & 0.00000 & 0.36568 & 10 & 55 & 0.00000\\ 
$CW\_L0.1\_S2.0 $ & 0.1 & 2.0 & 0.34822 & 0.00016 & 0.33953 & 10 & 53 & 0.00047\\
$CW\_L0.5\_S2.0 $ & 0.5 & 2.0 & 0.30815 & 0.01632 & 0.28657 & 10 & 62 & 0.05695\\
$CW\_L1.0\_S2.0 $ & 1.0 & 2.0 & 0.30582 & 0.07061 & 0.23215 & 10 & 70 & 0.30416\\
\hline
$CS\_L0.0\_S2.0 $ & 0.0 & 2.0 & 0.21836 & 0.00000 & 0.21731 & 10 & 36 & 0.00000 \\ 
$CS\_L0.1\_S2.0 $ & 0.1 & 2.0 & 0.21651 & 0.00008 & 0.21640 & 10 & 23 & 0.00037 \\
$CS\_L0.5\_S2.0 $ & 0.5 & 2.0 & 0.20916 & 0.00904 & 0.20159 & 10 & 32 & 0.04484 \\
$CS\_L1.0\_S2.0 $ & 1.0 & 2.0 & 0.24292 & 0.04255 & 0.20006 & 10 & 22 & 0.21269
\enddata
\tablenotetext{a-g}{Same as Table 1}
\end{deluxetable}

\end{document}